\documentclass[
  aps,
  prl,
  twocolumn,
  superscriptaddress,
  amsmath,
  amssymb,
  10pt
]{revtex4-2}

\usepackage{graphicx}
\usepackage{dcolumn}
\usepackage{bm}
\usepackage{physics}
\usepackage{braket}
\usepackage{mathrsfs}
\usepackage{color}
\usepackage[normalem]{ulem}
\usepackage{comment}
\usepackage{xcolor}
\usepackage{subfigure}
\usepackage{slashed}

\renewcommand{\pmat}[1]{\begin{pmatrix} #1 \end{pmatrix}}

\begin{document}

\title{Spontaneous Hawking Radiation from Dynamical Black-Hole Formation in Quantum Spin Systems}

\author{Shunichiro Kinoshita}
\email{kinoshita@neptune.kanazawa-it.ac.jp}
\affiliation{Mathematics, Science, Data Science and AI Program,
Kanazawa Institute of Technology,
Nonoichi, Ishikawa 921-8501, Japan}

\author{Keiju Murata}
\email{murata.keiju@nihon-u.ac.jp}
\affiliation{Department of Physics, College of Humanities and Sciences,
Nihon University, Sakura-josui,
Tokyo 156-8550, Japan}
\affiliation{Global R\&D Center for Business by Quantum-AI Technology (G-QuAT),
National Institute of Advanced Industrial Science and Technology (AIST),
Tsukuba, Ibaraki 305-8568, Japan}

\author{Daisuke Yamamoto}
\email{yamamoto.daisuke21@nihon-u.ac.jp}
\affiliation{Department of Physics, College of Humanities and Sciences,
Nihon University, Sakura-josui,
Tokyo 156-8550, Japan}
\affiliation{Global R\&D Center for Business by Quantum-AI Technology (G-QuAT),
National Institute of Advanced Industrial Science and Technology (AIST),
Tsukuba, Ibaraki 305-8568, Japan}
\affiliation{RIKEN Center for Quantum Computing (RQC),
Wako, Saitama 351-0198, Japan}

\author{Ryosuke Yoshii}
\email{ryoshii@rs.socu.ac.jp}
\affiliation{Center for Liberal Arts and Sciences,
Sanyo-Onoda City University,
Yamaguchi 756-0884, Japan}
\affiliation{International Institute for Sustainability with Knotted Chiral Meta Matter
(WPI-SKCM2), Hiroshima University,
Higashi-Hiroshima, Hiroshima 739-8526, Japan}

\begin{abstract}
We show that Hawking radiation spontaneously generated by dynamical
black-hole formation can be realized in a local quantum spin model.
Starting from the ground state corresponding to the early-time vacuum,
with no initially prepared excitations,
we follow the dynamics induced by the evolving geometry.
In the corresponding continuum quantum field theory, Hawking particle creation is
associated with nontrivial Bogoliubov mixing between the in- and
out-mode decompositions.
We derive the resulting renormalized local energy density
in the continuum quantum field theory, which corresponds to an energy-weighted
integral of the Hawking spectrum,
and identify its counterpart as a local spin observable whose expectation
value quantitatively reproduces the continuum prediction, including its
dependence on the surface gravity.
The agreement persists over a finite time window whose duration grows
logarithmically with system size, revealing the crossover from universal
continuum Hawking physics to lattice-cutoff dynamics.
These results establish a microscopic route to quantum simulation of
quantum field theory in curved spacetime, connecting field-theoretic particle creation
directly to experimentally accessible many-body observables.
\end{abstract}

\maketitle

\noindent
{\it Introduction.}---
Hawking radiation is one of the most striking predictions of quantum
field theory (QFT) in curved spacetime~\cite{Hawking:1975vcx}.
When a black hole forms, a quantum field initially in the vacuum state
gives rise to outgoing radiation with a thermal spectrum characterized
by a temperature set by the surface gravity, linking horizon geometry
to quantum theory and thermodynamics.
The associated black-hole evaporation raises the information-loss
problem~\cite{PhysRevD.14.2460,Mathur:2009hf}, which has motivated major
developments in quantum information and
holography~\cite{Page:1993df,Harlow:2014yka,Almheiri:2019psf,Penington:2019kki}.
At the same time, tracing Hawking quanta backward in time leads to an
exponential blueshift, raising the trans-Planckian question of how this
apparently universal low-energy prediction is affected by short-distance
physics~\cite{PhysRevD.44.1731,PhysRevD.51.2827,
Brandenberger:1999sw,Martin:2000xs}.
Yet Hawking radiation from astrophysical black holes has not been
directly observed, since their Hawking temperatures are extremely low
and the emission is overwhelmed by ambient radiation.
This has motivated efforts to reproduce and probe the underlying vacuum
particle-creation process in controllable laboratory systems.

Analogue-gravity systems have provided a major laboratory route to
horizon physics, in which excitations of a many-body medium are governed
by an effective field theory in a curved spacetime
~\cite{Unruh:1980cg,Visser:1997ux,Jacobson:1998ms,Barcelo:2005fc,
Jacquet:2020bar}.
Experiments with Bose--Einstein condensates have reported correlations
between Hawking and partner modes, thermal spectra consistent with the
expected Hawking temperature, and stationary emission
~\cite{Steinhauer:2015saa,MunozDeNova2019Hawking,
Kolobov2021Stationary}, although their interpretation remains under
discussion~\cite{Leonhardt:2016qdi}.
Complementary approaches using engineered lattice and quantum systems
have explored wave propagation, thermal signatures, stimulated emission,
particle-pair production, and detector responses in black-hole and
horizon settings~\cite{PhysRevResearch.2.023107,
PhysRevLett.130.016701,shi2023quantum,PhysRevB.109.014309,
Jaiswal:2026hawking}.
Beyond reproducing individual signatures, a broader goal is to establish
a quantitatively controlled quantum simulator of QFT in curved spacetime,
in which the target continuum theory is specified a priori, recovered in
a controlled continuum limit, and connected quantitatively to
microscopic observables.

Rapid advances in programmable quantum processors and simulators have
opened new possibilities for controlled studies of many-body dynamics
~\cite{Altman2021,Daley2022,Fauseweh2024}.
Across platforms ranging from superconducting circuits to trapped ions
and neutral-atom arrays, quantum systems now combine microscopic control
of state preparation and interactions with local readout
~\cite{Blais2020,Monroe2021,Browaeys2020,Ebadi2021}.
They also provide access to quantum-information properties, including
many-body correlations and entanglement, through flexible measurement
protocols~\cite{Elben2023,Hoke2023,Joshi2023,Karamlou2024}.
These capabilities have stimulated growing efforts to encode and
investigate quantum field theories on quantum devices
~\cite{Bauer2023NRP,Bauer2023PRXQ}.
Hawking radiation is a particularly compelling target: besides providing
a stringent benchmark for a curved-spacetime QFT simulator, a qubit-based
formulation would ultimately allow the radiation and its quantum
correlations to be interrogated within the same microscopic setting.

In this Letter, we demonstrate a microscopic spin realization of
spontaneous Hawking radiation generated by dynamical black-hole
formation.
Using an explicit QFT-to-spin dictionary developed in our previous
works~\cite{Kinoshita:2024ahu,Kinoshita:2026dhs}, we prescribe a spacetime
that evolves from an initially flat geometry into a black-hole geometry
and encode it directly in spatially and temporally varying spin couplings.
The spin system is initialized in the ground state corresponding to the
early-time vacuum, with no excitations seeded by hand.
As the geometry evolves, nontrivial Bogoliubov mixing arises between the
in- and out-mode decompositions, so that the initial vacuum contains
excitations with respect to the late-time modes.
We derive the resulting renormalized energy density in the continuum
QFT, identify its counterpart as a local spin observable, and show that
its expectation value quantitatively reproduces the QFT prediction,
including its dependence on the surface gravity.
We further identify a finite Hawking-radiation window whose duration
increases logarithmically with system size, revealing how universal
continuum Hawking physics gives way to lattice-cutoff dynamics. Together, these results provide a quantitative benchmark for microscopic
quantum simulation of QFT in curved spacetime and open a route to regimes
beyond analytically tractable free-field dynamics.
\\

\noindent
{\it Two-dimensional spacetime for dynamical black hole formation.}---
We consider a $(1+1)$-dimensional spacetime with the following metric:
\begin{equation}
ds^2 = -dt^2 + \left( dx - v(t,x)\, dt \right)^2\ ,
\label{metric}
\end{equation}
where the function $v(t,x)$ depends on both spatial and temporal coordinates.
The spatial coordinate is assumed to be defined on a finite interval $0 \leq x \leq \ell$. 
We set $\ell=\pi$ in the numerical simulations while keeping $\ell$ explicit in analytical expressions.

With an appropriate choice of $v(t,x)$, the metric describes a black-hole spacetime. 
Indeed, the null condition $ds^2=0$ gives
\begin{equation}
\frac{dx}{dt}=\pm1+v(t,x).
\end{equation}
The two signs correspond to right- and left-moving null rays, which can be viewed as being advected by a rightward ``wind'' $v(t,x)$. 
In a region with $v(t,x)>1$, both satisfy $dx/dt>0$ and propagate to the right, giving rise to a black-hole region from which causal curves cannot escape. This is precisely the defining feature of a black hole.
The determinant of the metric~\eqref{metric} is identically $-1$, and hence no coordinate singularity arises at the event horizon.
The Ricci scalar is
\begin{equation}
    R=\partial_x\bigl[2\partial_t v(t,x)
    +\partial_x\bigl(v^2(t,x)\bigr)\bigr]\ ,
\end{equation}
showing that no curvature singularity arises as long as $v(t,x)$ is smooth.

\begin{figure}
\begin{center}
\includegraphics[scale=0.27]{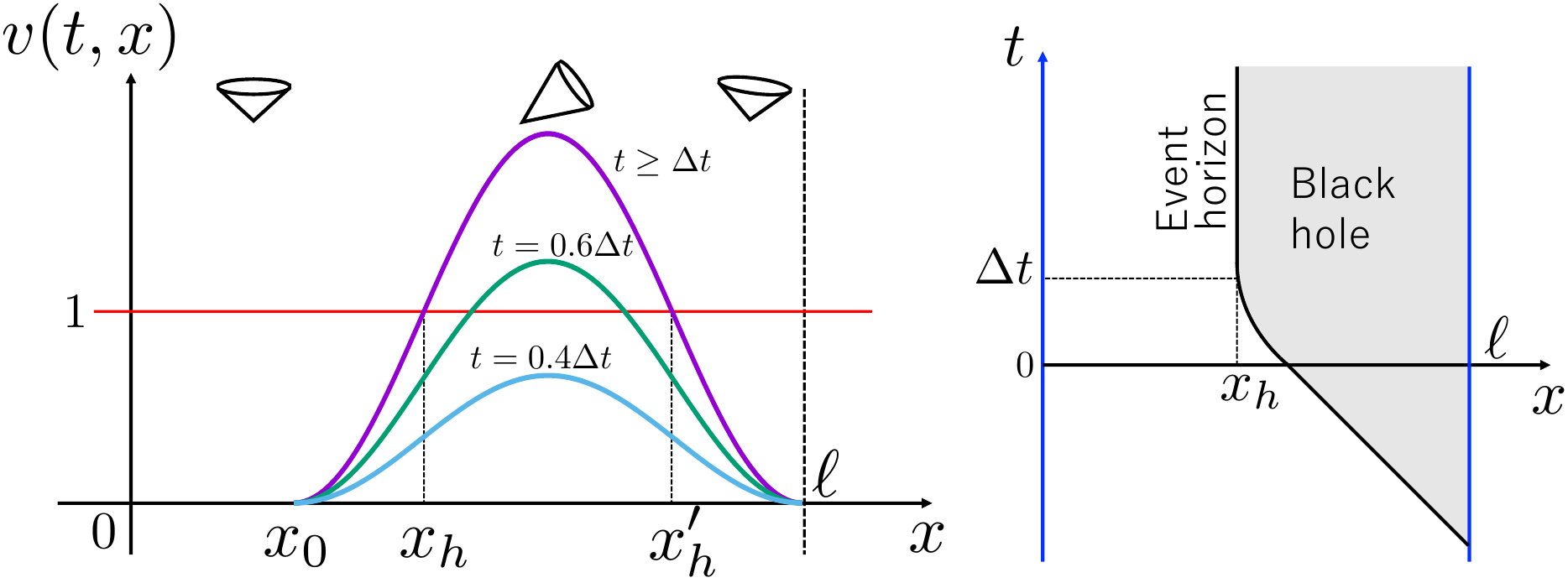}
\end{center}
\caption{(Left) Profile of $v(t,x)$ for $\ell=\pi$, $x_0=\pi/4$, and $\kappa=8/3$. (Right) Causal structure of the spacetime describing black hole formation.
The shaded region represents a black hole, from which the left boundary ($x=0$) cannot be reached. }
\label{v_and_h}
\end{figure}

We take $v(t,x)=h(t)v_0(x)$. For $x_0<x\leq\ell$, we choose
\begin{equation}
v_0(x)=
\left[1+\left(\frac{\kappa(\ell-x_0)}{2\pi}\right)^2\right]
\sin^2\left[\frac{\pi(x-x_0)}{\ell-x_0}\right],
\end{equation}
while $v_0(x)=0$ for $0\leq x\leq x_0$.
The time dependence is specified by
\begin{equation}
h(t)=\frac12\left[1-\cos\left(\frac{\pi t}{\Delta t}\right)\right]
\end{equation}
for $0<t<\Delta t$, with $h(t)=0$ for $t\leq0$ and
$h(t)=1$ for $t\geq\Delta t$.
The constant $\kappa >0$ in $v_0(x)$ will later be identified with the surface gravity of the black hole.
Figure~\ref{v_and_h} shows the spatial and temporal dependence of $v(t,x)$.
For $t \le 0$, the spacetime is static with $v(t,x)=0$, whereas for $t\ge \Delta t$, it becomes stationary with $v(t,x)=v_0(x)$.
The spacetime remains flat at all times in the region $x<x_0$.
We introduce this flat region so that Hawking radiation can be measured without contamination from local curvature effects.
The parameter $\Delta t$ characterizes the transition timescale.
To illustrate the propagation of null rays, the light cones for $t\geq \Delta t$ in each spatial region are also depicted in the left panel.
We denote by $x_h$ and $x_h'$ ($x_h<x_h'$) the two roots of $v_0(x)=1$.
The surface $x = x_h$ represents the location of the event horizon at the late time 
for observers in the left region $x < x_h$. 
Similarly, $x = x_h'$ corresponds to a white-hole horizon 
for observers in the right region $x > x_h'$.

The right panel of Fig.~\ref{v_and_h} schematically illustrates the causal structure of the spacetime.
For $t<\Delta t$, the event horizon is obtained by tracing null geodesics backward in time from $x=x_h$.
The white-hole horizon is omitted since it is not relevant to the present discussion.

For $t \geq \Delta t$, the surface gravity at the event horizon 
$x = x_h$ of this spacetime is given by
\begin{equation}
    \frac{dv_0(x)}{dx}\bigg|_{x=x_h}=\kappa=2\pi T_H\ ,
\end{equation}
where $T_H$ is the Hawking temperature. 
\\

\noindent
{\it Spin-system realization of QFT in a black-hole-forming spacetime.}---
As shown in our previous works~\cite{Kinoshita:2024ahu,Kinoshita:2025fij,Kinoshita:2026dhs}, Majorana fermions in a two-dimensional spacetime can be mapped onto a spin system. For the metric~(\ref{metric}), the corresponding open spin chain is
\begin{multline}
    H=-\frac{1}{2\varepsilon} \sum_{j=1}^{L}\bigg[
    \sigma^x_j \sigma^x_{j+1}\\
    - \frac{1}{2}v(t,x_{j+1/2}) (\sigma^x_j \sigma^y_{j+1}-\sigma^y_j \sigma^x_{j+1})
    +  \sigma^z_j \bigg]\ .
    \label{IsingforQFT}
\end{multline}
To avoid splitting the sum, we set operators at site $L+1$ to zero. We assign a coordinate $x_j$ to each site and define $x_{j+1/2}$ for the link between sites $j$ and $j+1$ by
\begin{equation}
x_j=\varepsilon (j-1)\ ,\quad x_{j+1/2}=x_j+\frac{\varepsilon}{2}\ ,\quad \varepsilon = \frac{\ell}{L}\ .
\end{equation}
This spin system is a transverse-field Ising model with a Dzyaloshinskii--Moriya interaction (TI-DM model).
The QFT on a finite interval requires boundary conditions, which fix otherwise free boundary parameters in the spin system~\cite{Kinoshita:2026dhs}. 
The Hamiltonian above incorporates this choice; details are provided in the Supplemental Material.

In the flat region $x<x_0$, where $v(t,x)=0$, the spin Hamiltonian reduces to the transverse-field Ising model, and we define the local energy density as
\begin{equation}
    \mathcal{H}_j\equiv -\frac{1}{2\varepsilon^2} \bigg[
    \frac{1}{2}(\sigma^x_j \sigma^x_{j+1}+\sigma^x_{j-1} \sigma^x_{j})
    +  \sigma^z_j \bigg] .
    \label{Hdensitysigma}
\end{equation}
In defining the Hamiltonian density, the link terms are averaged over the bonds to the left and right of site $j$ so that their contributions are treated symmetrically. In contrast, the on-site term $\sigma^z_j$ is kept unchanged.

Although no regularization is required in the discrete system, the energy density contains a large contribution that diverges as $\varepsilon\to0$, corresponding to the ultraviolet divergence in the continuum limit. We remove it by subtracting the initial energy density.
We first prepare the state $|\Omega\rangle$ as the ground state of the Hamiltonian~(\ref{IsingforQFT}) with $v(t,x)=0$, namely, the ground state of the transverse-field Ising model. 
Taking $|\Omega\rangle$ as the initial state at $t=0$, we then evolve it with the Hamiltonian~(\ref{IsingforQFT}) with time-dependent $v(t,x)$ and denote the resulting state by $|\psi(t)\rangle$. 
The renormalized energy density is defined as
\begin{equation}
\rho_j(t)
\equiv
\langle \psi(t)|\mathcal{H}_j|\psi(t)\rangle
-
\langle \Omega|\mathcal{H}_j|\Omega\rangle \, .
\label{Erenom}
\end{equation}
Details of the numerical calculation are provided in the End Matter.

In QFT, the late-time renormalized energy density in the flat region $x<x_0$ can be calculated analytically as
\begin{equation}
\rho_\textrm{QFT}(t,x)\to
\frac{1}{48\pi}\left(\kappa^2+\frac{\pi^2}{\ell^2}\right),
\qquad (t\to\infty).
\label{EhawkingTot}
\end{equation}
A detailed derivation is provided in the Supplemental Material. 
The $\kappa$-dependent term can also be understood directly from the
thermal Fermi--Dirac distribution. For a massless Majorana fermion,
a single chiral sector has the thermal energy density
\begin{equation}
\int_0^\infty \frac{d\omega}{2\pi}
\frac{\omega}{e^{\omega/T_H}+1}
=\frac{\pi}{24}T_H^2 .
\end{equation}
Since the late-time flat region contains both left- and right-moving
components, the total thermal contribution is
$\rho_{\rm th}=\pi T_H^2/12=\kappa^2/(48\pi)$, corresponding to the
$(1+1)$-dimensional Stefan--Boltzmann law for Fermi-Dirac statistics.
The remaining $\pi/(48\ell^2)$ term is the finite-size Casimir
contribution.

We now examine whether the spin system reproduces this QFT prediction. 
Although the Hawking spectrum contains more detailed information, the corresponding particle-number operator is highly nonlocal in the spin representation~\cite{Kinoshita:2024ahu}. 
We therefore focus on the energy density, which is an integrated measure of the spectrum and remains a local observable in the spin system, making it more accessible experimentally.
\\

\begin{figure}
\begin{center}
\includegraphics[scale=0.6]{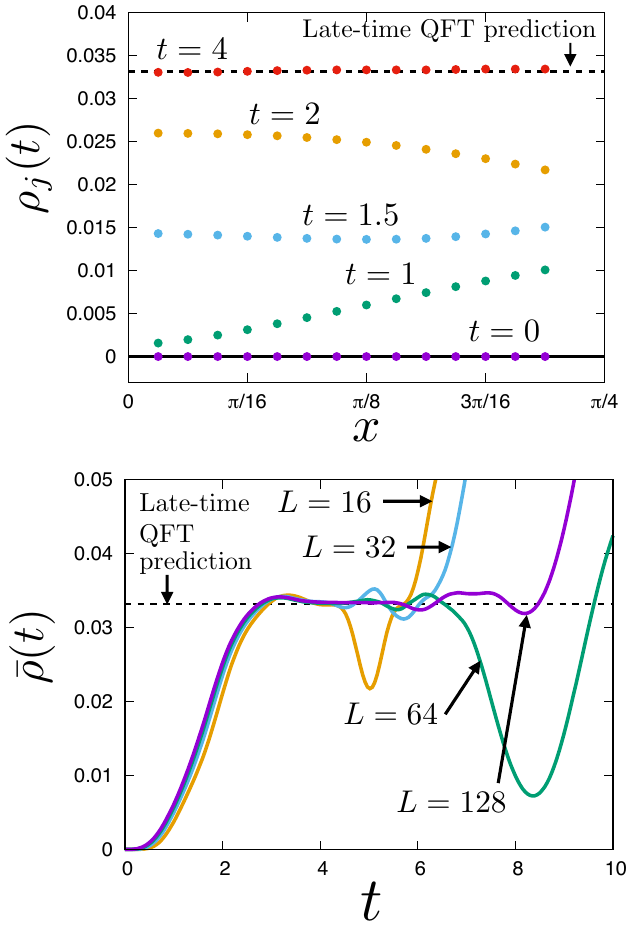}
\end{center}
\caption{
(Top) Energy density computed in the spin system. Only the flat region $x<x_0-\varepsilon$ is shown. The black dashed line indicates the QFT prediction~(\ref{EhawkingTot}).
(Bottom) Spatially averaged energy density over the flat region as a function of time for various numbers of sites $L$; the black dashed line indicates the QFT prediction.
The parameters are $\ell=\pi$, $x_0=\pi/4$, $\kappa=2$, and $\Delta t=2$. The number of sites is $L=64$ in the top panel and $L=16,32,64,$ and $128$ in the bottom panel.
}
\label{Hfig}
\end{figure}

\noindent
{\it Energy density of Hawking radiation in the spin system.}---
The resulting renormalized energy density~(\ref{Erenom}) is shown in the top panel of Fig.~\ref{Hfig} at $t=0,1,1.5,2,$ and $4$.
The energy densities are displayed only for sites satisfying $x_j<x_0-\varepsilon$ (ensuring that the neighboring site $j+1$ also lies in the flat region).
The energy density is initially zero but increases as time evolves. After some time, it temporarily relaxes toward the QFT prediction~(\ref{EhawkingTot}).
Although not shown in the top panel, at later times the energy density deviates from the QFT prediction. This deviation is clearly visible in the bottom panel, as discussed below.

To characterize the time dependence, we spatially average the energy density over the $N$ sites in the flat region:
\begin{equation}
    \bar{\rho}(t) \equiv 
    \frac{1}{N}\sum_{x_j<x_0-\varepsilon}\rho_j(t)\ .
\end{equation}
The bottom panel of Fig.~\ref{Hfig} shows that, for all values of $L$, $\bar{\rho}(t)$ develops a plateau consistent with the QFT prediction~(\ref{EhawkingTot}), indicating that the flat region is filled with Hawking radiation after the black hole has formed.
At late times, however, the plateau disappears and $\bar{\rho}(t)$ deviates from the QFT prediction as the finite lattice scale becomes relevant, reflecting the trans-Planckian problem discussed in the next section.
The plateau duration increases with $L$. 
Remarkably, even for $L=16$, where the flat region contains only two sites, the QFT prediction is reproduced over a finite time window, $2.5\lesssim t\lesssim4$.

\begin{figure}
\begin{center}
\includegraphics[scale=0.6]{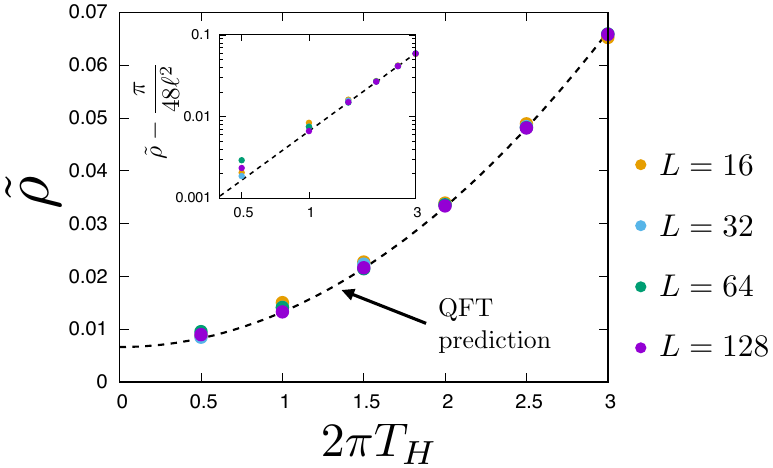}
\end{center}
\caption{
Plateau height of the spatially averaged energy density as a function of $\kappa=2\pi T_H$.
The dots show the results for the discrete spin system, while the black dashed line indicates the QFT prediction~(\ref{EhawkingTot}).
The parameters are $\ell=\pi$, $x_0=\pi/4$, and $\Delta t=2$, with the number of sites $L=16,32,64,$ and $128$.
The inset shows the same data on a log-log scale after subtracting the constant term $\pi/(48\ell^2)$ from the QFT prediction.
}
\label{Hplateau}
\end{figure}

To quantify the plateau height, we define the time-averaged energy density
\begin{equation}
    \tilde{\rho}
    =\frac{1}{t_2-t_1}\int_{t_1}^{t_2}dt\,\bar{\rho}(t)\ .
\end{equation}
We choose $t_1=\Delta t+c_1/\kappa$ and $t_2=\Delta t+c_2(\log L)/\kappa$, with $c_1=2$ and $c_2=1.5$, to exclude the initial transient and late-time deviation, respectively.
The choices of $t_1$ and $t_2$ are motivated in the next section in connection with the trans-Planckian problem.
Figure~\ref{Hplateau} shows $\tilde{\rho}$ as a function of $\kappa=2\pi T_H$ for various $L$, in good agreement with the quadratic QFT prediction~(\ref{EhawkingTot}). 
The inset shows the same results on a log-log scale after subtracting the Casimir contribution $\pi/(48\ell^2)$. 
(A quantitative analysis of the convergence with $L$ is provided in the Supplemental Material.)
Remarkably, good agreement with QFT is obtained even for small systems with $L=16$ and $32$, despite their shorter plateau durations. 
This demonstrates that Hawking radiation can be quantitatively characterized through the energy density of relatively small spin systems, suggesting its experimental accessibility.

Additional results for various $\kappa$ and $\Delta t$ are provided in the Supplemental Material. 
For $\Delta t=0$, strong oscillations obscure the plateau, suggesting that a sufficiently smooth metric evolution facilitates experimental observation of Hawking radiation.
\\

\noindent
{\it Trans-Planckian problem.—}
The late-time deviation can be understood in terms of the trans-Planckian problem. 
In QFT, tracing Hawking radiation backward in time leads to an exponential blueshift to arbitrarily short wavelengths. 
In our spin system, the lattice spacing $\varepsilon=\ell/L$ provides the shortest length scale, giving rise to an analogue of the trans-Planckian problem~\cite{PhysRevD.44.1731,PhysRevD.51.2827,
Brandenberger:1999sw,Martin:2000xs}.

This timescale can be estimated by considering a wave packet of initial width $\sim\varepsilon$ near the horizon at $t=\Delta t$.
Its width grows as $\varepsilon e^{\kappa(t-\Delta t)}$ due to the exponential redshift and becomes comparable to $\ell$ at
\begin{equation}
t_{\rm T.P.}-\Delta t
\sim \frac{1}{\kappa}\log\frac{\ell}{\varepsilon}
\sim \frac{1}{\kappa}\log L .
\label{tTP}
\end{equation}
This explains why the late-time deviation in Fig.~\ref{Hfig} is delayed as $L$ increases.

Stationary Hawking radiation requires an equilibration time $t_{\rm eq}-\Delta t\sim\kappa^{-1}$.
Thus, for $t_{\rm eq}\lesssim t\lesssim t_{\rm T.P.}$, the spin system reproduces the stationary Hawking radiation predicted by QFT, explaining both the plateau in Fig.~\ref{Hfig} and its extension with increasing $L$.
\\

\noindent
{\it Conclusion}---
We have shown that a local quantum spin system encoding dynamical
black-hole formation develops a renormalized local energy density that
quantitatively reproduces the continuum-QFT prediction for Hawking
radiation, including its dependence on the surface gravity.
The evolution starts from the ground state corresponding to the
early-time vacuum, with no initially prepared excitations.
The agreement persists over a finite time window, while the onset of
lattice-induced deviations is delayed logarithmically with system size.
This behavior identifies the lattice cutoff not merely as a limitation
of the simulation, but as an explicit microscopic scale that exposes
the crossover from continuum Hawking physics to short-distance dynamics.

More broadly, evaluating the same renormalized local energy density in
the continuum theory and in the spin system makes the
continuum-to-microscopic correspondence directly testable.
The same construction specifies the ingredients for implementation on
programmable quantum platforms: state preparation, time-dependent
couplings, and local measurements.
Extensions beyond free fields and vacuum initial conditions could bring
interacting and far-from-equilibrium QFTs in curved spacetime within
reach of quantum simulation, including strongly correlated regimes
relevant to holography.
From this perspective, the present Hawking-radiation setting points
toward a broader program of using microscopic quantum systems to examine
how universal predictions of curved-spacetime QFT survive, or are
modified by, short-distance quantum dynamics.

\begin{acknowledgments}
We are grateful to the members of QuEra Computing and G-QuAT at AIST for many stimulating discussions. This work was supported by JSPS KAKENHI Grant Numbers~ JP23K25830 (DY), JP24K06890 (DY), JP26K00664 (DY), and JST PRESTO Grant No.~JPMJPR245D (DY).
\end{acknowledgments}

\bibliography{ref}

\clearpage
\appendix
\onecolumngrid
\begin{center}
{\bfseries END MATTER}
\end{center}
\twocolumngrid

{\it Numerical calculation in the spin system.}---
Here, we describe the numerical procedure used for the spin system. We work in the Heisenberg picture throughout the calculation. We first apply the Jordan--Wigner transformation to the spin operators,
\begin{equation}
\sigma^+_j
=
\prod_{l=1}^{j-1}(1-2c_l^\dagger c_l)c_j,
\qquad
\sigma_j^z=1-2c_j^\dagger c_j\ ,
\label{JW}
\end{equation}
where $\sigma_j^\pm \equiv (\sigma^x_j \pm i\sigma^y_j)/2$.
We further introduce the Hermitian operators $a_j^\pm$ through
\begin{equation}
c_j
=
\frac{1}{2}
\left[
a_j^+-a_j^-
-i(a_j^++a_j^-)
\right],
\end{equation}
which satisfy
\begin{equation}
\{a_j^s,a_{j'}^{s'}\}
=
\delta_{jj'}\delta_{ss'}.
\end{equation}
Applying these transformations to the spin Hamiltonian~(\ref{IsingforQFT}), we obtain a quadratic Hamiltonian in terms of $a_j^\pm$,
\begin{multline}
H
=
-\frac{i}{2\varepsilon}
\sum_{j=1}^{L}
\Big[
(1+v_{j+1/2})a_j^+a_{j+1}^+\\
-(1-v_{j+1/2})a_j^-a_{j+1}^-\\
+a_j^-a_{j+1}^+
-a_j^+a_{j+1}^-
+2a_j^+a_j^-
\Big],
\label{Hbyapm}
\end{multline}
where $v_{j+1/2}=v(t,x_{j+1/2})$.
This Hamiltonian  can be written in the following form using a $2L \times 2L$ block matrix $M(t)$ as
\begin{equation}
    H=\frac{1}{2}\pmat{\bm{a}^+{}^T & \bm{a}^-{}^T}
M(t) \pmat{\bm{a}^+ \\ \bm{a}^-}\ ,
\end{equation}
where $\bm{a}^\pm$ denote the $L$-component column vectors defined by $\bm{a}^\pm = (a_1^\pm,\cdots,a_L^\pm)^T$.
The $2L \times 2L$ matrix $M(t)$ is given by 
\begin{equation}
M(t)=\frac{i}{2\varepsilon}
\begin{pmatrix}
\tilde{A}(t)-A & S\\
-S & \tilde{A}(t)+A
\end{pmatrix},
\end{equation}
where the $L\times L$ matrices $\widetilde{A}(t)$, $A$, and $S$ are defined by
\begin{align}
\tilde{A}_{jk}(t)
&=
v_{j-1/2}(t)\delta_{j,k+1}
-v_{j+1/2}(t)\delta_{j+1,k},\\
A_{jk}
&=
\delta_{j+1,k}-\delta_{j,k+1},\\
S_{jk}
&=
\delta_{j+1,k}+\delta_{j,k+1}-2\delta_{jk},
\end{align}
where $v_{j+1/2}(t)\equiv v(t,x_{j+1/2})$.
Thus, $M(t)^T=-M(t)$ and $M(t)^\dagger=M(t)$.

The Heisenberg equation for $\bm{a}^\pm$ is
\begin{equation}
    \frac{d}{dt}\pmat{\bm{a}^+ \\ \bm{a}^-}=-iM(t) \pmat{\bm{a}^+ \\ \bm{a}^-}\ .
\end{equation}
We expand the operators as
\begin{equation}
    \pmat{\bm{a}^+ \\ \bm{a}^-} = \sum_{n=1}^{L} (\vec{\phi}_n(t) \gamma_n + \vec{\phi}_n(t)^\ast  \gamma_n^\dagger)\ ,
\end{equation}
where the mode function $\vec{\phi}_n(t)$ is a $2L$-component complex column vector whose components are written as
\begin{equation}
    \vec{\phi}_n(t)=(\phi^+_{n1}(t), \cdots, \phi^+_{nL}(t),\phi^-_{n1}(t),\cdots,\phi^-_{nL}(t))^T\ .
\end{equation}
The mode functions $\vec{\phi}_n$ for $n=1,\ldots,L$ satisfy 
\begin{equation}
    \frac{d}{dt}\vec{\phi}_n=-iM(t) \vec{\phi}_n\ .
    \label{phieqspin}
\end{equation}
Since the equations above conserve the standard inner product, we require the mode functions to be orthonormal, as 
\begin{equation}
    \vec{\phi}_n(t)\cdot \vec{\phi}_m(t)=0\ ,\quad \vec{\phi}_n(t)^\ast \cdot \vec{\phi}_m(t)=\delta_{nm}\ .
\end{equation}
With this normalization condition, the operators $\gamma_n$ satisfy the canonical anticommutation relations: 
\begin{equation}
    \{\gamma_n,\gamma_m^\dagger\}=\delta_{nm}\ ,\quad \{\gamma_n,\gamma_m\}=\{\gamma_n^\dagger,\gamma_m^\dagger\}=0\ .
\end{equation}

Let us consider the case $t<0$. In this case, the matrix 
$M_0 \equiv M(t)\vert_{t<0}$ is time-independent. 
Since $M_0$ is an antisymmetric Hermitian matrix, its eigenvalues necessarily appear in pairs of real numbers with opposite signs. 
We denote them as
\begin{equation}
\omega_1,\omega_2,\ldots,\omega_L,\,-\omega_1,-\omega_2,\ldots,-\omega_L ,
\end{equation}
where $0<\omega_1<\omega_2<\cdots<\omega_L$. Let $\vec v_n$ be the normalized eigenvector corresponding to the eigenvalue $\omega_n$. Then the eigenvector corresponding to $-\omega_n$ is given by $\vec{v}_n^\ast$.
Using these eigenvalues and eigenvectors of $M_0$, the mode functions for $t<0$ can be written as
\begin{equation}
\vec{\phi}_n(t)=\vec{v}_n e^{-i\omega_n t}\ .
\end{equation}
The Hamiltonian for $t<0$ is diagonalized by these mode functions as
\begin{equation}
H|_{t<0}=\sum_{n=1}^L \omega_n \gamma_n^\dagger \gamma_n + E_0 \ ,
\end{equation}
where $E_0=-\sum_{n=1}^L \omega_n/2$. Therefore, the ground state $|\Omega\rangle$ for $t<0$ is defined as the state satisfying
\begin{equation}
\gamma_n|\Omega\rangle = 0 \,, \quad (\forall n\in\{1,2,\cdots,L\}) .
\end{equation}

For $t\geq0$, the matrix $M(t)$ becomes time-dependent, and the mode functions are obtained by numerically solving the differential equation~(\ref{phieqspin}) using the fourth-order Runge--Kutta method.
The initial condition for $\vec{\phi}_n(t)$ is set to $\vec{\phi}_n(0)=\vec{v}_n$. This ensures continuity with the positive-frequency mode functions for $t<0$. 

Once the mode functions are obtained, the energy density can be evaluated directly from them. In the flat region, the energy density~(\ref{Hdensitysigma}) can be rewritten in terms of the operators $a_j^\pm$ as
\begin{multline}
\mathcal{H}_j=-\frac{i}{4\varepsilon^2}\Big[a_j^+ a_{j+1}^+ + a_{j-1}^+ a_{j}^+
-a_j^- a_{j+1}^--a_{j-1}^- a_{j}^-\\
a_j^- a_{j+1}^+ - a_j^+ a_{j+1}^- + a_{j-1}^- a_{j}^+ - a_{j-1}^+ a_{j}^-
+4\,a_j^+ a_j^-
\Big]\ .
\label{Hpmint}
\end{multline}
To evaluate the vacuum expectation value of the energy density, it is sufficient to compute the equal-time Wightman functions $\langle a_j^s(t)a_{j'}^{s'}(t) \rangle$ where $s,s'=+,-$ and $\langle \cdots \rangle = \langle\Omega| \cdots|\Omega \rangle$.  Using the mode expansion, these correlation functions can be expressed in terms of the mode functions as
\begin{equation}
\langle a_j^s(t) a_{j'}^{s'}(t')\rangle 
=\sum_{n=1}^L\phi^s_{nj}(t) \phi^{s'}_{nj'}(t')^\ast\ ,
\end{equation}
Substituting these correlation functions into the expression for the energy density above, we obtain the time-dependent energy density used in the numerical analysis.

\end{document}


\title{Supplemental Material for\\[3pt]
``Simulating Hawking radiation arising from dynamical black hole formation
in quantum spin systems''}

\author{Shunichiro Kinoshita}
\email{kinoshita@neptune.kanazawa-it.ac.jp}
\affiliation{Mathematics, Science, Data Science and AI Program,
Kanazawa Institute of Technology,
Nonoichi, Ishikawa 921-8501, Japan}

\author{Keiju Murata}
\email{murata.keiju@nihon-u.ac.jp}
\affiliation{Department of Physics, College of Humanities and Sciences,
Nihon University, Sakura-josui,
Tokyo 156-8550, Japan}
\affiliation{Global R\&D Center for Business by Quantum-AI Technology (G-QuAT),
National Institute of Advanced Industrial Science and Technology (AIST),
Tsukuba, Ibaraki 305-8568, Japan}

\author{Daisuke Yamamoto}
\email{yamamoto.daisuke21@nihon-u.ac.jp}
\affiliation{Department of Physics, College of Humanities and Sciences,
Nihon University, Sakura-josui,
Tokyo 156-8550, Japan}
\affiliation{Global R\&D Center for Business by Quantum-AI Technology (G-QuAT),
National Institute of Advanced Industrial Science and Technology (AIST),
Tsukuba, Ibaraki 305-8568, Japan}
\affiliation{RIKEN Center for Quantum Computing (RQC),
Wako, Saitama 351-0198, Japan}

\author{Ryosuke Yoshii}
\email{ryoshii@rs.socu.ac.jp}
\affiliation{Center for Liberal Arts and Sciences,
Sanyo-Onoda City University,
Yamaguchi 756-0884, Japan}
\affiliation{International Institute for Sustainability with Knotted Chiral Meta Matter
(WPI-SKCM2), Hiroshima University,
Higashi-Hiroshima, Hiroshima 739-8526, Japan}

\begin{abstract}
This Supplemental Material provides details of the theoretical and numerical analyses supporting the results presented in the main text. We first formulate a massless Majorana field in the black-hole-forming spacetime and analytically derive the late-time mode functions, Wightman functions, and energy density of Hawking radiation. We then present the mapping to the corresponding quantum spin system and demonstrate explicitly that its Hamiltonian density reproduces the continuum field theory. Finally, we provide additional numerical results, including the decomposition of the energy density into right- and left-moving contributions, its dependence on the surface gravity and black-hole formation timescale, convergence toward the continuum limit, and the dynamics of the mode functions.
\end{abstract}

\maketitle

\section{Majorana fermions in a spacetime with black hole formation}
\subsection{Lagrangian and Hamiltonian}

We consider the spacetime metric introduced in the main text,
\begin{equation}
ds^2 = -dt^2 + \left( dx - v(t,x)\,dt \right)^2 ,
\label{metric}
\end{equation}
where the spatial coordinate $x$ is defined on the interval $0\leq x\leq \ell$.
We adopt the following separable form for $v(t,x)$:
\begin{equation}
v(t,x)=h(t)v_0(x),
\label{vtimesh}
\end{equation}
where
\begin{equation}
v_0(x)=
\begin{cases}
0, & (0\leq x \leq x_0),\\
\displaystyle
\left[1+\left(\frac{\kappa(\ell-x_0)}{2\pi}\right)^2\right]
\sin^2\left[\frac{\pi(x-x_0)}{\ell-x_0}\right],
& (x_0<x\leq\ell),
\end{cases}
\end{equation}
and
\begin{equation}
h(t)=
\begin{cases}
0, & (t\leq0),\\
\displaystyle
\frac{1}{2}\left[1-\cos\left(\frac{\pi t}{\Delta t}\right)\right],
& (0<t<\Delta t),\\[2mm]
1, & (t\geq\Delta t).
\end{cases}
\end{equation}
Let $x_h$ and $x_h'$ ($x_h<x_h'$) denote the two roots of $v_0(x)=1$.
For the present choice of $v_0(x)$, they are explicitly given by
\begin{equation}
x_h
=x_0+\frac{\ell-x_0}{\pi}
\arctan\left[\frac{2\pi}{\kappa(\ell-x_0)}\right],
\qquad
x_h'
=\ell-\frac{\ell-x_0}{\pi}
\arctan\left[\frac{2\pi}{\kappa(\ell-x_0)}\right].
\end{equation}
At late times, $x=x_h$ and $x=x_h'$ correspond to the event and white-hole horizons, respectively. The surface gravity at the event horizon is given by $\kappa=dv_0(x)/dx|_{x=x_h}$. 

To formulate the fermion theory in this curved spacetime, we introduce the zweibein
\begin{equation}
e^0=e^0_{\ \mu}dx^\mu=dt,
\qquad
e^1=e^1_{\ \mu}dx^\mu=dx-v(t,x)\,dt.
\label{zwei1}
\end{equation}
The corresponding dual vectors are
\begin{equation}
e_0=e_0^{\ \mu}\partial_\mu
=\partial_t+v(t,x)\,\partial_x,
\qquad
e_1=e_1^{\ \mu}\partial_\mu
=\partial_x.
\label{zwei2}
\end{equation}
They satisfy
\begin{equation}
g_{\mu\nu}e_a^{\ \mu}e_b^{\ \nu}=\eta_{ab},
\qquad
g^{\mu\nu}e^a_{\ \mu}e^b_{\ \nu}=\eta^{ab},
\end{equation}
where $\eta_{ab}=\eta^{ab}=\mathrm{diag}(-1,1)$.
Following Ref.~\cite{Kinoshita:2024ahu}, we consider massless Majorana fermions in the $(1+1)$-dimensional black-hole-forming spacetime described by Eq.~\eqref{metric}, which can be mapped onto a quantum spin system.

The Lagrangian for massless Majorana fermions in curved spacetime is given by~\cite{Polyakov:1981re,deLacroix:2023uem,deLacroix:2023uem,Kinoshita:2024ahu}\footnote{
For a Majorana fermion in $(1+1)$ dimensions, the spin connection term does not appear in the Lagrangian. Therefore, the derivatives in the action can be written as ordinary derivatives rather than covariant ones.
}
\begin{equation}
    \mathcal{L}=
    -\frac{i}{4}\sqrt{-g}\,\bar{\psi} (\Slash{\partial}-\ldiff{\Slash{\partial}}) \psi\ , \quad
    \psi=\pmat{a^+ \\ a^-}\ .
\end{equation}
where $a^\pm$ are real (Hermitian) fermionic operators satisfying $(a^\pm)^\dagger = a^\pm$. 
We define the Dirac conjugate by $\bar{\psi} = \psi^T \Gamma^0$, 
and Dirac operators in curved spacetime by
$\slashed{\partial}\psi = \Gamma^a e_a^{\ \mu} \partial_\mu \psi$ and 
$\bar{\psi}\overleftarrow{\Slash{\partial}}=e_a^{\ \mu}\partial_\mu \bar{\psi} \Gamma^a$.
We adopt the Weyl representation for the Clifford algebra:\footnote{
In our previous works~\cite{Kinoshita:2024ahu,Kinoshita:2025fij,Kinoshita:2026dhs}, we employed the Majorana representation. 
Let $\psi_W$ and $\psi_M$ denote the spinors in the Weyl and Majorana representations, respectively. Their relation is given by $\psi_W = 2^{-1/2}(\sigma^x+\sigma^z)\psi_M$.
}
\begin{equation}
    \Gamma^0 = -i \sigma^y=\pmat{0 & -1\\1&0}\ ,\quad \Gamma^1=\sigma^x=\pmat{0 & 1\\1&0}\ .
\end{equation}
They satisfy $\{\Gamma^a,\Gamma^b\}=2\eta^{ab}$.
In this representation, the chiral matrix is diagonal, with $\Gamma^3 \equiv \Gamma^0 \Gamma^1 = - \sigma^z$. For the zweibeins defined in Eqs.~(\ref{zwei1}) and (\ref{zwei2}), 
the Lagrangian is explicitly written as
\begin{equation}
    \mathcal{L}=\frac{i}{4}[ a^+ \lrdiff{\partial}_t a^+ + (1+v(t,x)) a^+\lrdiff{\partial}_x a^+ 
    +a^- \lrdiff{\partial}_t  a^- -(1-v(t,x)) a^-\lrdiff{\partial}_x a^-]\ ,
\end{equation}
where $a^s\lrdiff{\partial}_\mu a^s = a^s \partial_\mu a^s -(\partial_\mu a^s) a^s$ ($s=\pm$). 
The Hamiltonian density corresponding to the above Lagrangian is given by
\begin{equation}
    \mathcal{H}=-\frac{i}{4}[(1+v(t,x)) a^+ \lrdiff{\partial}_x a^+
    -(1-v(t,x))a^- \lrdiff{\partial}_x a^-]\ ,
    \label{HQFT}
\end{equation}
Here, $a^\pm$ satisfy the canonical equal-time anticommutation relations: 
$\{a^+(t,x), a^+(t,y)\}=\{a^-(t,x), a^-(t,y)\}=\delta(x-y)$ and $\{a^+(t,x), a^-(t,y)\}=0$.  
Since the left- and right-moving sectors of the above Hamiltonian density decouple, the Hamiltonian density can be decomposed into the contributions for each mode, defined as 
\begin{equation}
\mathcal{H}^+\equiv -\frac{i}{4}(1+v(t,x)) a^+ \lrdiff{\partial}_x a^+\ ,\quad 
    \mathcal{H}^-\equiv \frac{i}{4}(1-v(t,x))a^- \lrdiff{\partial}_x a^-\ .
    \label{Hpm}
\end{equation}
In this section, we will evaluate the expectation values of $\mathcal{H}^\pm$ generated by the Hawking radiation produced during black-hole formation.

\subsection{Boundary conditions}
Since we are considering a field theory defined on a finite interval with boundaries at $x=0,\ell$, we must specify appropriate boundary conditions there. 
As discussed in Refs.~\cite{Witten:2015aoa,Hashimoto:2016kxm,Al-Hashimi:2016vfd,Kinoshita:2026dhs}, the allowed boundary conditions for a Majorana field are determined by the requirement that the inner product be time-independent. 
In the present case, the spacetime is flat in the vicinity of the boundaries, and the admissible boundary conditions are either 
$a^+ + a^- = 0$ or $a^+ - a^- = 0$, as shown in Ref.~\cite{Kinoshita:2026dhs}. 
(Physically, upon reflection at the boundary, the right- and left-moving modes must be exchanged with at most a sign.)
Which of these boundary conditions is chosen does not significantly affect Hawking radiation and is therefore not essential for the discussion in this paper. 
Accordingly, we choose the following boundary condition so that the corresponding spin system becomes simpler: 
\begin{equation}
   (a^+ + a^-)\big|_{x=0} = 0 \, , \qquad
(a^+ - a^-)\big|_{x=\ell} = 0 \, .
\label{bc}
\end{equation}
With this choice of boundary condition, a parameter in the corresponding spin system can be fixed to the constant value $1$. This point will be discussed in section~\ref{spinHamiltonian}.

\subsection{Equations of motion and mode functions}

The equations of motion for $a^\pm$ are given by
\begin{equation}
\bigg[
\partial_t 
+ 
\begin{pmatrix}
1+v(t,x) & 0 \\
0 & -(1-v(t,x))
\end{pmatrix}
\partial_x 
+ \frac{v'(t,x)}{2}
\bigg]
\begin{pmatrix}
a^+ \\
a^-
\end{pmatrix}
= 0 \, ,
\label{apmeq}
\end{equation}
where the prime denotes differentiation with respect to $x$.
We expand the operators $a^\pm$ as
\begin{equation}
\begin{pmatrix}
a^+(t,x) \\
a^-(t,x)
\end{pmatrix}
=
\sum_n
\left(
\vec{\phi}_n(t,x)\, \gamma_n
+
\vec{\phi}_n^{\,*}(t,x)\, \gamma_n^\dagger
\right) ,
\label{aexpand}
\end{equation}
where $\vec{\phi}_n$ is a two-component vector,
\begin{equation}
\vec{\phi}_n(t,x)
=
\begin{pmatrix}
\phi_n^+(t,x) \\
\phi_n^-(t,x)
\end{pmatrix},
\end{equation}
whose components are $c$-number functions. The mode functions satisfy
\begin{equation}
\bigg[
\partial_t 
+ 
\begin{pmatrix}
1+v(t,x) & 0 \\
0 & -(1-v(t,x))
\end{pmatrix}
\partial_x 
+ \frac{v'(t,x)}{2}
\bigg]
\vec{\phi}_n
= 0 \, .
\label{phieq}
\end{equation}
The boundary condition (\ref{bc}) implies that the mode functions must also satisfy 
\begin{equation}
   (\phi_n^+ + \phi_n^-)\big|_{x=0} = 0 \, , \qquad
(\phi_n^+ - \phi_n^-)\big|_{x=\ell} = 0 \, .
\label{bcphi}
\end{equation}
We impose the orthonormality conditions for mode functions as
\begin{equation}
\int_0^\ell dx\, 
\vec{\phi}_n \cdot \vec{\phi}_m = 0 \, , 
\qquad
\int_0^\ell dx\, 
\vec{\phi}_n \cdot \vec{\phi}_m^{\,*} = \delta_{nm} \, .
\end{equation}
It can be verified from Eq.~(\ref{phieq}) that the inner product defined above is time-independent.
Then, one can check that the operators $\gamma_n$ satisfy the canonical anticommutation relations:
\begin{equation}
\{\gamma_n,\gamma_m\}=0 \, , 
\qquad 
\{\gamma_n,\gamma_m^\dagger\}=\delta_{nm} \, .
\end{equation}

\subsection{Solutions for the mode functions}

At early times $t\leq 0$, we have $v(t,x)=0$, and therefore the mode functions can be easily solved as
\begin{equation}
    \begin{pmatrix}
\phi_n^+ \\
\phi_n^-
\end{pmatrix}=\frac{1}{\sqrt{2\ell}}\pmat{e^{-i\omega_n(t-x)}\\-e^{-i\omega_n(t+x)}}\ ,
    \label{phisolv0}
\end{equation}
with
\begin{equation}
    \omega_n=\frac{\pi}{\ell}(n-\frac{1}{2})\ ,\quad n=1,2,\ldots\ .
\end{equation}
One can verify that these mode functions satisfy the boundary conditions~(\ref{bcphi}).

At general times, the mode functions cannot be obtained in a simple closed form.
However, at sufficiently late times, $\kappa(t-\Delta t)\gg1$, they take a simple asymptotic form in the flat region $x<x_0$.
In the following, we derive this late-time expression.
Here we study the late-time behavior of the solutions to the Majorana fermion equation of motion (\ref{apmeq}) or (\ref{phieq}). 
In order to analyze the solutions, it is convenient to rewrite the metric (\ref{metric}) in double-null coordinates. 
We introduce functions $U(t,x)$ and $V(t,x)$ satisfying the transport equations
\begin{equation}
(\partial_t + (1+v(t,x))\partial_x)\,U(t,x) = 0, 
\qquad
(\partial_t - (1-v(t,x))\partial_x)\,V(t,x) = 0 .
\label{UVeq}
\end{equation}
Using these functions, we perform a coordinate transformation $U=U(t,x)$ and $V=V(t,x)$, under which the metric can be written as
\begin{equation}
ds^2=\frac{dU\,dV}{U'V'} \, ,
\end{equation}
where the prime denotes differentiation with respect to $x$.
In order to obtain explicit solutions of Eq.~(\ref{UVeq}), we must specify initial and boundary conditions. 
For $t \le 0$, we take the usual double-null coordinates in flat spacetime,
$U = t - x$, $V = t + x$.
We also assume that the boundary condition $U=V$ holds at $x=0$.
With these initial and boundary conditions specified, the solution for $U$ and $V$ is uniquely determined. 
(The boundary at $x=\ell$ lies inside the event horizon, and therefore does not affect the analysis as long as we restrict attention to the region outside the horizon.)

\begin{figure}
\begin{center}
\includegraphics[scale=0.6]{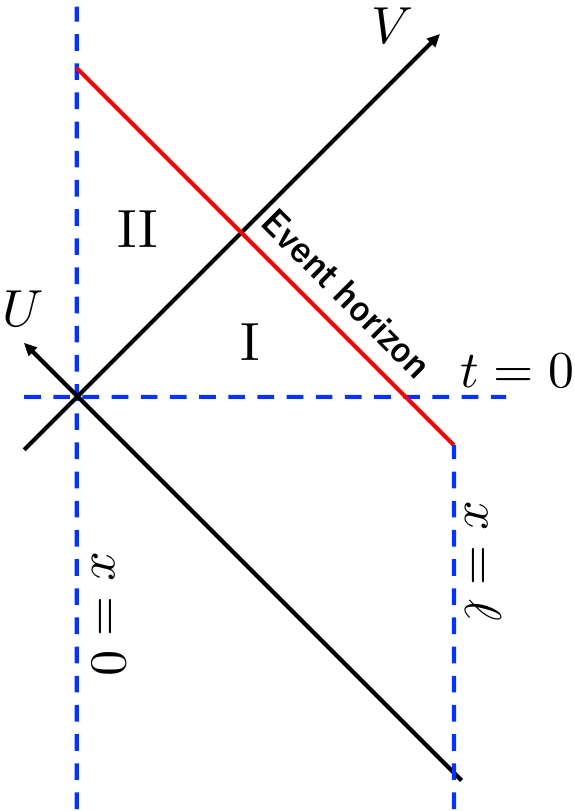}
\end{center}
\caption{Spacetime structure in $(U,V)$-coordinates.
}
\label{UVdiagram}
\end{figure}

Figure~\ref{UVdiagram} shows the spacetime structure in the $U,V$ coordinates. 
The curves of constant $V$ correspond to left-moving null rays,
while $U$ is constant along right-moving null rays.
The initial surface $t=0$ and the boundary $x=0$ correspond to $U=-V$ and $U=V$, respectively. 
Since the event horizon is a null surface, it is given by a surface of constant $V$.

Since the theory of a massless Majorana fermion in $(1+1)$ dimensions is a CFT, 
the equation of motion should reduce to the wave equation in flat spacetime 
after an appropriate rescaling of the fields. 
Using Eq.~(\ref{UVeq}) to rewrite Eq.~(\ref{phieq})
in the $(U,V)$ coordinates,
one finds that the equations simplify to
\begin{equation}
    \partial_V \left(\frac{\phi^+_n}{\sqrt{-U'}}\right)=0\ ,\quad \partial_U \left(\frac{\phi^-_n}{\sqrt{V'}}\right)=0\ .
\end{equation}
Therefore, the general solution to the equation of motion can be written in terms of arbitrary functions $F(U)$ and $G(V)$ as
\begin{equation}
\phi^+_n = \sqrt{-U'}\,F(U), 
\qquad 
\phi^-_n = \sqrt{V'}\,G(V) \, .
\end{equation}

The initial conditions for the mode functions are given by Eq.~(\ref{phisolv0}), which implies
\begin{equation}
F(-x)=\frac{1}{\sqrt{2\ell}}e^{i\omega_n x}, \qquad 
G(x)=-\frac{1}{\sqrt{2\ell}}e^{-i\omega_n x}.
\end{equation}
Furthermore, from the boundary condition~(\ref{bcphi}), we obtain 
\begin{equation}
\sqrt{-U'|_{x=0}}\,F(V)
=
-\sqrt{V'|_{x=0}}\,G(V).
\end{equation}
Therefore, the mode functions can be written as
\begin{equation}
\phi^+_n =
\begin{cases}
    \sqrt{\frac{-U'}{2\ell}}\, e^{-i\omega_n U}. & (U \leq 0)\\
    \sqrt{\frac{-U'}{2\ell}}\sqrt{\frac{V'}{-U'}\big|_{x=0}}\, e^{-i\omega_n U} & (U>0)
\end{cases}
\ ,\quad
    \phi^-_n = -\sqrt{\frac{V'}{2\ell}}\,e^{-i\omega_n V}.
\end{equation}

From the discussion so far, it is clear that once the solution of Eq.~(\ref{UVeq}) is known, the mode functions are immediately determined by the expressions above. 
We now derive the late-time form of the null coordinate $V$,
from which the late-time behaviour of the mode functions follows.
To analyze the solution for $V$, let us consider the equation satisfied by the characteristic curves,
\begin{equation}
\frac{dx}{dt} = -1 + v(t,x) \, .
\label{chareq}
\end{equation}
This is nothing but the geodesic equation for the left-moving null rays.
Imposing the initial condition $x=x_{\rm ini}$ at $t=0$, we denote the solution of the characteristic curve by $x=X(t;x_{\rm ini})$. 
Along this curve one finds
\begin{equation}
\frac{d}{dt} V\!\left(t, X(t;x_{\rm ini})\right)=0 \, ,
\end{equation}
which shows that $V(t,x)$ is constant along the characteristics. 
In particular, since the initial condition for $V$ is $V(0,x)=x$, we obtain
\begin{equation}
V\!\left(t, X(t;x_{\rm ini})\right)=x_{\rm ini} \, .
\end{equation}
Therefore, if we solve $x=X(t;x_{\rm ini})$ for $x_{\rm ini}$ and write the result as $x_{\rm ini}=x_{\rm ini}(t,x)$, we have
\begin{equation}
V(t,x)=x_{\rm ini}(t,x) \, .
\label{xiniisV}
\end{equation}
Thus, once the family of geodesics is known, the function $V$ can be determined.

Let us consider sufficiently late times. 
A left-moving null geodesic that reaches a spacetime point at late times propagates close to the horizon around at $t=0$. 
Let $x=X_h(t)$ denote the null geodesic that lies exactly on the horizon. 
We write a nearby geodesic of $x=X_h(t)$ as
\begin{equation}
X(t)=X_h(t)+\delta X(t) .
\end{equation}
Expanding the geodesic equation to linear order in the deviation, we obtain
\begin{equation}
\frac{d\delta X(t)}{dt}
=
\frac{\partial v}{\partial x}(t,X_h(t))\,\delta X(t) .
\end{equation}
Therefore, at $t=\Delta t$, we obtain
\begin{equation}
\delta X(\Delta t)
=\alpha \delta X(0)\ ,\quad 
\alpha = \exp\!\left[
\int_0^{\Delta t}
\partial_x v(t,X_h(t))\,dt
\right]\ .
\label{linsol}
\end{equation}
For $t\geq \Delta t$, the spacetime becomes stationary, so that $v(t,x)=v_0(x)$. 
In this case the solution of the geodesic equation~(\ref{chareq}) is given by
\begin{equation}
t-\Delta t = - x_\ast(x) + c ,\qquad 
x_\ast(x)=\int_0^x \frac{dx'}{1-v_0(x')} \, ,
\end{equation}
where $c$ is an integration constant.
Near the horizon $x=x_h$, the "tortoise coordinate" $x_\ast(x)$ behaves as
\begin{equation}
x_\ast(x) \simeq -\frac{1}{\kappa}\log [\kappa (x_h-x)] +d\ ,
\end{equation}
where 
\begin{equation}
d = \lim_{x\to x_h-0} \{x_\ast(x)+\frac{1}{\kappa}\log [\kappa (x_h-x)]\} = x_0 - \frac{1}{\kappa}
\log\!\left(
\frac{1}{4}+\frac{\pi^2}{\kappa^2 (\ell - x_0)^2}
\right)
\end{equation}
is the subleading term in the series expansion.
On the other hand, in the flat region $x<x_0$, we have $v_0(x)=0$, and therefore
$x_\ast(x)=x$.
The constant $c$ is fixed by matching the geodesic
to the linearized solution  (\ref{linsol}) at $t=\Delta t$ as
\begin{equation}
c=x_\ast\!\left(x_h+\delta X(\Delta t)\right)\simeq -\frac{1}{\kappa}\log (-\kappa\delta X(\Delta t))+d \ .
\end{equation}
Using Eq.~(\ref{linsol}) together with $\delta X(0)=x_{\rm ini}-X_h(0)$, we obtain
\begin{equation}
x_{\rm ini} = X_h(0) - \frac{1}{\kappa \alpha} e^{-\kappa(t-\Delta t-d + x_\ast(x))}\ .
\end{equation}
Therefore, from Eq.(\ref{xiniisV}), the late-time expression for the function $V$ in the flat region $x<x_0$ is given by
\begin{equation}
    V(t,x)= X_h(0)-e^{-\kappa(t-t_0+x)}\ ,
\end{equation}
where we have introduced the constant $t_0=\Delta t +d-\log(\kappa\alpha)/\kappa$, 
but its explicit form are not important for the computation of the Wightman functions as in the main text.

For $t \ge \Delta t$ and in the flat region $x<x_0$, the equation for $U$ is $(\partial_t + \partial_x)U=0$, whose general solution is $U=H(t-x)$, where $H$ is an arbitrary function.
To determine the late-time behavior of the function $U$, we consider region II in Fig.~\ref{UVdiagram}. 
In this region, the functional form of $U$ is fixed by the boundary condition at $x=0$.
From the boundary condition $U|_{x=0}=V|_{x=0}$ at $x=0$, we have $H(t)=V(t,0)=X_h(0)-e^{-\kappa (t-t_0)}$.
This gives the late-time solution for $U$ in the flat region $x<x_0$ as
\begin{equation}
U(t,x)=X_h(0)-e^{-\kappa(t-t_0-x)} .
\end{equation}

Ignoring the irrelevant constant phase factor $e^{i\omega_n X_h(0)}$, the mode functions at late times in the flat region $(x<x_0)$ take the form
\begin{equation}
\begin{pmatrix}
\phi_n^+ \\
\phi_n^-
\end{pmatrix}=
\sqrt{\frac{\kappa}{2\ell}}
\begin{pmatrix}
e^{-\kappa (t-t_0-x)/2}\exp[i\omega_n e^{-\kappa(t-t_0-x)}] \\
-\,e^{-\kappa (t-t_0+x)/2}\exp[i\omega_n e^{-\kappa(t-t_0+x)}]
\end{pmatrix}\ .
\label{phisollate}
\end{equation}

\subsection{Wightman functions}

The initial vacuum $|\Omega\rangle$ is defined by
\begin{equation}
\gamma_n |\Omega\rangle = 0 ,\qquad (\forall n \in \{1,2,\ldots\}) .
\label{vacdef}
\end{equation}
One can verify that this state indeed corresponds to the ground state for $t\leq 0$. In the following, we compute the Wightman functions of $a^\pm$ in this state.  As will be shown later, physical quantities such as the energy density can be derived from these Wightman functions.
From Eq.~(\ref{aexpand}) and  the definition of the vacuum~(\ref{vacdef}), the Wightman functions are given by
\begin{equation}
    \langle a^s(t,x) a^{s'}(t',x') \rangle = \sum_{n=1}^\infty \phi_n^s(t,x) \{\phi_n^{s'}(t',x')\}^\ast \ ,
    \label{aaphiphi}
\end{equation}
where $s,s'=+, -$ and $\langle \cdots \rangle \equiv \langle \Omega|\cdots|\Omega\rangle$.

First, let us consider the case of flat spacetime, $v(t,x)=0$. In this case, the mode functions are given by Eq.~(\ref{phisolv0}) for all times. The sum over $n$ in Eq.~(\ref{aaphiphi}) forms an infinite geometric series and can therefore be evaluated analytically, yielding
\begin{equation}
    \langle a^s(t,x) a^{s'}(t',x') \rangle|_{v=0} =
    \frac{ss'}{4i \ell} \frac{1}{\sin[\frac{\pi}{2\ell}\{(t-sx)-(t'-s'x')\}]}\ ,
    \label{Wfuncv0}
\end{equation}
where, to ensure convergence of the infinite series, we interpret the time $t$ as being shifted infinitesimally into the lower half of the complex plane as $t\to t-i0$.

Next, let us consider the case where $v(t,x)$ gives a black-hole-forming spacetime as in Eq.~(\ref{vtimesh}). 
In this case, the mode functions at late times are given by Eq.~(\ref{phisollate}). 
The sum over $n$ again forms an infinite geometric series and can therefore be evaluated analytically as
\begin{equation}
\begin{split}
     &\langle a^s(t,x) a^{s'}(t',x') \rangle 
     = -\frac{ss'\kappa}{4i\ell} 
    \frac{e^{-\kappa (t-t_0-sx+t'-t_0-s'x')/2}}{\sin[\frac{\pi}{2\ell}\{e^{-\kappa(t-t_0-sx)}-e^{-\kappa(t'-t_0-s'x')}\}]}\\
    &\simeq -\frac{ss'\kappa}{2\pi i} 
    \frac{e^{-\kappa (t-sx+t'-s'x')/2}}{e^{-\kappa(t-sx)}-e^{-\kappa(t'-s'x')}}
    =\frac{ss'\kappa }{4\pi i} \frac{1}{\sinh[\kappa\{(t-sx)-(t'-s'x')\}/2]}\ .
\end{split}
\label{Wfunclate}
\end{equation}
In the second equality, we have approximated $\sin$ by a linear function, using the fact that $t$ and $t'$ are sufficiently large.

\subsection{Energy density of Hawking radiation}

In the flat region $x<x_0$, the energy densities of the left- and right-moving modes are given by Eq.~(\ref{Hpm}) as
\begin{equation}
\langle \mathcal{H}^\pm(t,x) \rangle 
    = \mp \frac{i}{4} \langle a^\pm(t,x) \lrdiff{\partial}_x a^\pm(t,x) \rangle 
    =\mp \frac{i}{4} (\partial_{x'}-\partial_{x})  \langle a^\pm(t,x) a^\pm(t,x') \rangle \big |_{x'\to x}\ .
\end{equation}
The last expression shows that the energy density can be evaluated from the Wightman functions by applying the coincident-point limit. 
However, since these Wightman functions~(\ref{Wfunclate}) diverge in the coincident-point limit, the energy density above also diverges. 
This divergence is a UV divergence characteristic of quantum field theory. 
It can be regularized by introducing a reference Hamiltonian $H_\text{ref}$ and considering the difference between the expectation value of an operator evolved with $H$ and that with $H_\text{ref}$. 
Here, we adopt the Hamiltonian with $v(t,x)=0$ as the reference Hamiltonian, i.e., $H_\text{ref}=H|_{v=0}$.
Using Eqs.~(\ref{Wfuncv0}) and (\ref{Wfunclate}), and subtracting the coincident-point divergence of the Wightman functions, the energy density is obtained as
\begin{align}
    \rho^\pm(t,x)&\equiv 
    \langle \mathcal{H}^\pm(t,x) \rangle - \langle \mathcal{H}^\pm(t,x) \rangle|_{v=0}
    = \frac{1}{96\pi}\left(\kappa^2 + \frac{\pi^2}{\ell^2}\right)\ ,\label{Ehawking_pm}\\
    \rho(t,x)&\equiv \rho^+(t,x) + \rho^-(t,x)=\frac{1}{48\pi}\left(\kappa^2 + \frac{\pi^2}{\ell^2}\right)\ .
    \label{EhawkingTot}
\end{align}
The $\kappa$-independent terms in the above expressions for the energy density correspond to the Casimir energy in flat spacetime on the interval $0 \leq x \leq \ell$, with the opposite sign. The first term corresponds to the energy density of the Hawking radiation.
At late times in the flat region, the Hawking radiation equilibrates and the energy density approaches the constant value obtained above. Our proposal is to measure this quantity using the corresponding spin system.


\section{Corresponding spin system}
\label{Spinsection}

\subsection{Hamiltonian}
\label{spinHamiltonian}

The general spin Hamiltonian corresponding to the Majorana field theory in the metric~(\ref{metric}) is given by~\cite{Kinoshita:2024ahu,Kinoshita:2025fij,Kinoshita:2026dhs}
\begin{multline}
    H=-\frac{1}{4\varepsilon} \sum_{j=1}^L\bigg\{
    \left(\cos\zeta(t,x_{j+1/2}) + p(t,x_{j+1/2})\right)\sigma^x_j \sigma^x_{j+1}
    - \left( \cos\zeta(t,x_{j+1/2}) - p(t,x_{j+1/2})\right)\sigma^y_j \sigma^y_{j+1} \\
    - \left(v(t,x_{j+1/2}) + \sin\zeta(t,x_{j+1/2})\right) \sigma^x_j \sigma^y_{j+1} 
    + \left(v(t,x_{j+1/2}) -  \sin\zeta(t,x_{j+1/2})\right) \sigma^y_j \sigma^x_{j+1}\\
    + \big[2p(t,x_j) - \varepsilon \big(\partial_x p(t,x_j)
   - \partial_t\zeta(t,x_j) - v(t,x_j)\partial_x\zeta(t,x_j)\big)\big] \sigma^z_j \bigg\} ,
    \label{IsingGeneral}
\end{multline}
and operators at site $L+1$ are defined to be zero.
Here,
\begin{equation}
x_j=\varepsilon(j-1),\qquad
x_{j+1/2}=x_j+\frac{\varepsilon}{2},\qquad
\varepsilon=\frac{\ell}{L}.
\end{equation}
The Hamiltonian~(\ref{IsingGeneral}) contains two arbitrary functions, $p(t,x)$ and $\zeta(t,x)$, as discussed in Ref.~\cite{Kinoshita:2024ahu}.
The function $\zeta(t,x)$ corresponds to a local phase rotation of the fermion field and thus simply reflects the freedom of field redefinition in the continuum theory.
On the other hand, $p(t,x)$ contributes only to subleading terms in the continuum limit and therefore does not affect the resulting field theory as $L\to\infty$. Nevertheless, $p(t,x)$ plays the role of a Wilson term that suppresses fermion doublers, and it is therefore preferable to choose $p(t,x)\neq0$~\cite{Kinoshita:2026dhs}.
For a field theory on a finite interval, the boundary values of $p(t,x)$ determine the boundary conditions of the Majorana field~\cite{Kinoshita:2026dhs}.
The boundary conditions in Eq.~(\ref{bc}) correspond to
\begin{equation}
p(t,0)=p(t,\ell)=1.
\end{equation}
We therefore choose the simplest function satisfying these conditions,
\begin{equation}
p(t,x)=1,
\qquad
\zeta(t,x)=0\ .
\end{equation}
With these choices, the Hamiltonian reduces to
\begin{equation}
    H=-\frac{1}{2\varepsilon} \sum_{j=1}^{L}\bigg[
    \sigma^x_j \sigma^x_{j+1}
    - \frac{1}{2}v(t,x_{j+1/2}) (\sigma^x_j \sigma^y_{j+1}-\sigma^y_j \sigma^x_{j+1})
    +  \sigma^z_j \bigg]\ .
    \label{IsingforQFT}
\end{equation}

\subsection{Continuum limit}

We now demonstrate explicitly that the local operators in the spin system reproduce the QFT Hamiltonian densities in the continuum limit.
We first map the spin operators to fermionic operators using the Jordan--Wigner transformation,
\begin{equation}
\sigma^+_j=\prod_{l=1}^{j-1}(1-2c_l^\dagger c_l) c_j\ ,\quad 
    \sigma^-_j=\prod_{l=1}^{j-1}(1-2c_l^\dagger c_l) c_j^\dagger\ ,\quad 
    \sigma_j^z=1-2c_j^\dagger c_j\ ,
    \label{JW}
\end{equation}
which introduces fermionic operators $c_j$ satisfying the canonical anticommutation relations, $\{c_j,c_{j'}^\dagger\}=\delta_{jj'}$ and $\{c_j,c_{j'}\}=\{c_j^\dagger,c_{j'}^\dagger\}=0$.
In terms of these fermionic operators, the Hamiltonian of the spin system can be written as
\begin{equation}
 H=- \frac{1}{2\varepsilon} \sum_{j=1}^{L}\bigg\{ 
 c_{j+1}c_j +c_j^\dagger c_{j+1}^\dagger
 + c_j^\dagger c_{j+1}+ c_{j+1}^\dagger c_j
 + iv_{j+1/2}( c_j^\dagger c_{j+1}- c_{j+1}^\dagger c_j)+
1-2c_j^\dagger c_j\bigg\} ,
 \label{IsingforQFTc}
\end{equation}
where $v_{j+1/2}=v(t,x_{j+1/2})$. 
We introduce Hermitian operators $a_j^\pm$ through
\begin{equation}
c_j=\frac{1}{2}(a^+_j-a^-_j - i(a^+_j+a^-_j))\ ,\quad
    c_j^\dagger=\frac{1}{2}(a^+_j-a^-_j + i(a^+_j+a^-_j))\ ,
\end{equation}
which satisfy $\{a^s_j,a^{s'}_{j'}\}=\delta_{jj'}\delta_{ss'}$. 
Substituting these expressions into the Hamiltonian, we obtain
\begin{equation}
H
=-\frac{i}{2\varepsilon}\sum_{j=1}^{L}\Big[
(1+v_{j+1/2})\,a_j^+ a_{j+1}^+
-(1-v_{j+1/2})\,a_j^- a_{j+1}^-
+a_j^- a_{j+1}^+ - a_j^+ a_{j+1}^-
+2\,a_j^+ a_j^-
\Big].
\label{Hbyapm}
\end{equation}
We define the Hamiltonian density in the spin system as
\begin{equation}
\mathcal{H}_j=\mathcal{H}^+_j+ \mathcal{H}^-_j + \mathcal{H}^\text{int}_j\ ,
\label{Hdensityspin}
\end{equation}
where 
\begin{equation}
\begin{split}
&\mathcal{H}_j^+=-\frac{i}{4\varepsilon^2}\Big[(1+v_{j+1/2})\,a_j^+ a_{j+1}^+ + (1+v_{j-1/2})\,a_{j-1}^+ a_{j}^+]\ ,\\
&\mathcal{H}_j^-=\frac{i}{4\varepsilon^2}\Big[(1-v_{j+1/2})\,a_j^- a_{j+1}^-+(1-v_{j-1/2})\,a_{j-1}^- a_{j}^-\Big]\ ,\\
&\mathcal{H}^\text{int}_j=-\frac{i}{4\varepsilon^2}\Big[a_j^- a_{j+1}^+ - a_j^+ a_{j+1}^- + a_{j-1}^- a_{j}^+ - a_{j-1}^+ a_{j}^-
+4\,a_j^+ a_j^-
\Big]\ .
\end{split}
\label{Hpmint}
\end{equation}
Here, $\mathcal{H}^+_j$ and $\mathcal{H}^-_j$ correspond to the Hamiltonian densities of the right and left movers, respectively. 
The term $\mathcal{H}^\text{int}_j$ represents their interaction, and it will be shown later that it vanishes in the continuum limit.
In defining the Hamiltonian densities at site $j$ as above, the link terms are averaged so that the contributions from the left and right of site $j$ are treated symmetrically. 
In contrast, the bond term $a_j^+ a_j^-$ is kept as it is.
Using the Hamiltonian density defined above the Hamiltonian is written as
\begin{equation}
    H=\varepsilon \sum_{j=2}^{L-1} \mathcal{H}_j + B_1+B_L\ .
\end{equation}
where boundary terms are given by
\begin{equation}
\begin{split}
    B_1&=-\frac{i}{4\varepsilon}[
    (1+v_{3/2})\,a_1^+ a_{2}^+
-(1-v_{3/2})\,a_1^- a_{2}^-+a_1^- a_{2}^+ - a_1^+ a_{2}^-+4a_1^+ a_1^-]\ .\\
B_L&=-\frac{i}{4\varepsilon}[
    (1+v_{L-1/2})\,a_{L-1}^+ a_{L}^+
-(1-v_{L-1/2})\,a_{L-1}^- a_{L}^-+a_{L-1}^- a_{L}^+ - a_{L-1}^+ a_{L}^-+4a_L^+ a_L^-]\ .
\end{split}
\end{equation}

We now consider the continuum limit of this theory. We introduce continuum fields by
\begin{equation}
    a^\pm(x_j)=\frac{a^\pm_j}{\sqrt{\varepsilon}}\ .
\end{equation}
Noting the relation $2 a_j^\pm a_{j+1}^\pm = a_j^\pm (a_{j+1}^\pm-a_j^\pm) - (a_{j+1}^\pm-a_j^\pm)a_j^\pm$, the continuum limit of $\mathcal{H}_j^\pm$ follows immediately and is given by
\begin{equation}
\begin{split}
&\mathcal{H}_j^+ 
\to 
-\frac{i}{4} (1+v(t,x))\,a^+ \lrdiff{\partial}_x a^+
\ ,\\
&\mathcal{H}_j^- \to 
\frac{i}{4} (1-v(t,x))\,a^- \lrdiff{\partial}_x a^-\ ,\quad (\varepsilon\to 0)\ .
\end{split}
\label{Hpmconti}
\end{equation}
They are identical to Eq.~(\ref{Hpm}). 
For $\mathcal{H}^\text{int}_j$, noting that the expression in the bracket can be written as $a_j^-(a_{j+1}^+-2a_j^++a_{j-1}^+)-a_j^+(a_{j+1}^--2a_j^-+a_{j-1}^-)$, the continuum limit can be taken straightforwardly, yielding
\begin{equation}
    \mathcal{H}^\text{int}_j \simeq -\frac{i\varepsilon}{4}
    (a^-\partial_x^2 a^+-a^+\partial_x^2 a^-)\to 0\ ,\quad
    (\varepsilon\to 0)\ .
\end{equation}
Therefore, the Hamiltonian density defined in Eq.~(\ref{Hdensityspin}) coincides with that of the QFT in the continuum limit. Similary, in the continuum limit, boundary terms are given by 
\begin{equation}
    B_1\to \frac{i}{2}a^- a^+|_{x=0}\ ,\quad 
    B_L\to \frac{i}{2}a^- a^+|_{x=\ell}\ ,\quad (\varepsilon\to 0)\ .
\end{equation}
Due to the boundary condition (\ref{bc}), these terms all reduce to the square of the same fermionic operator. 
By the equal-time anticommutation relations, they become (divergent) constants and are therefore irrelevant to the dynamics of the quantum field.

\subsection{Hamiltonian density in the spin-operator representation}

We can rewrite the Hamiltonian densities~(\ref{Hpmint}) in terms of the original spin operators as
\begin{equation}
    \begin{split}
        \mathcal{H}_j^\pm&=-\frac{1}{16\varepsilon^2}[
(1\pm v_{j+1/2})(\sigma^x_j\pm\sigma^y_j)(\sigma^x_{j+1}\mp\sigma^y_{j+1})
+(1\pm v_{j-1/2})(\sigma^x_{j-1}\pm\sigma^y_{j-1})(\sigma^x_{j}\mp\sigma^y_{j})
        ]\ ,\\
\mathcal{H}^{\text{int}}_j
&=
-\frac{1}{8\varepsilon^2}
\Big[
\sigma_j^x\sigma_{j+1}^x+\sigma_j^y\sigma_{j+1}^y
+\sigma_{j-1}^x\sigma_j^x+\sigma_{j-1}^y\sigma_j^y
+4\sigma_j^z
\Big] .
    \end{split}
\end{equation}
The sum of the three terms above corresponds to the summand of Eq.~(\ref{Hdensityspin}), where the link terms are averaged between the left and right sites, and can be written as
\begin{equation}
    \mathcal{H}_j=-\frac{1}{2\varepsilon^2} \bigg[
    \frac{1}{2}(\sigma^x_j \sigma^x_{j+1}+\sigma^x_{j-1} \sigma^x_{j})
    - \frac{1}{4}(v_{j+1/2} (\sigma^x_j \sigma^y_{j+1}-\sigma^y_j \sigma^x_{j+1})+v_{j-1/2}(\sigma^x_{j-1}\sigma^y_{j}-\sigma^y_{j-1} \sigma^x_{j}))
    +  \sigma^z_j \bigg] .
    \label{Hdensitysigma}
\end{equation}
Since we measure the energy density of Hawking radiation in the flat region, where $v(t,x)=0$, we can set $v_{j\pm1/2}=0$ in the above expressions.

\subsection{Renormalized energy density}
In the discrete system, the expectation values of the Hamiltonian density defined above are of course finite, so that regularization is unnecessary.
However, it contains a ``large'' contribution that corresponds to the UV divergence in the continuum limit (i.e.\ a quantity that diverges in the limit $\varepsilon \to 0$). 
To subtract this large contribution, we follow the same procedure as in the QFT case [see Eqs.~(\ref{Ehawking_pm}) and (\ref{EhawkingTot})]. On the spin system side, we likewise consider the difference from the expectation value at $v(t,x)=0$. We define the vacuum for $v(t,x)=0$ as $|\Omega\rangle$ and denote expectation values by $\langle \cdots \rangle \equiv \langle \Omega | \cdots | \Omega \rangle$. 
The renormalized energy density is then defined by 
\begin{equation}
\begin{split}
    &\rho_j^\pm(t)
=
\langle \mathcal{H}^\pm_j \rangle
-
\left. 
\langle \mathcal{H}^\pm_j \rangle
\right|_{v=0}\ ,\quad
\rho_j^\text{int}(t)
=
\langle \mathcal{H}^\text{int}_j \rangle
-
\left. 
\langle \mathcal{H}^\text{int}_j \rangle
\right|_{v=0}\ ,\\
&\rho_j(t)=\langle \mathcal{H}_j \rangle
-
\left. 
\langle \mathcal{H}_j \rangle
\right|_{v=0}
=\rho_j^+(t)+\rho_j^-(t)+\rho_j^\text{int}(t)\ .
\end{split}
\label{rhospin_def}
\end{equation}
Since the system size $\ell$ is finite in the initial state, this procedure effectively subtracts the contribution of the Casimir energy. 
It also removes, as a by-product, boundary-induced contributions that are universally present, but irrelevant to the Hawking radiation in the bulk region.


\section{Detailed analysis of Hawking radiation in the spin system}
\label{resultsection}

\subsection{Decomposition of the energy density}
\begin{figure}
  \centering
\subfigure[Energy density of right-moving modes]
 {\includegraphics[scale=0.7]{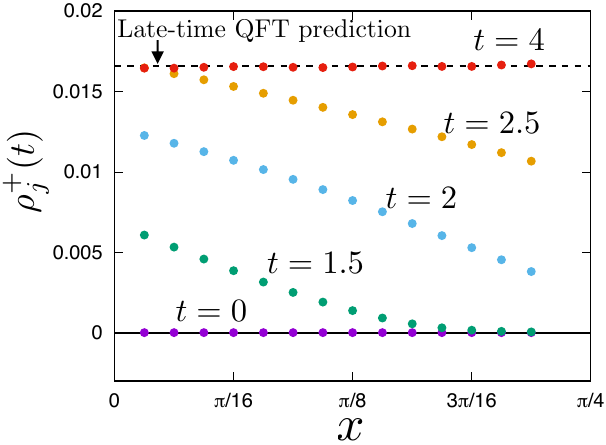}\label{Hpfig}
  }
  \subfigure[Energy density of left-moving modes]
 {\includegraphics[scale=0.7]{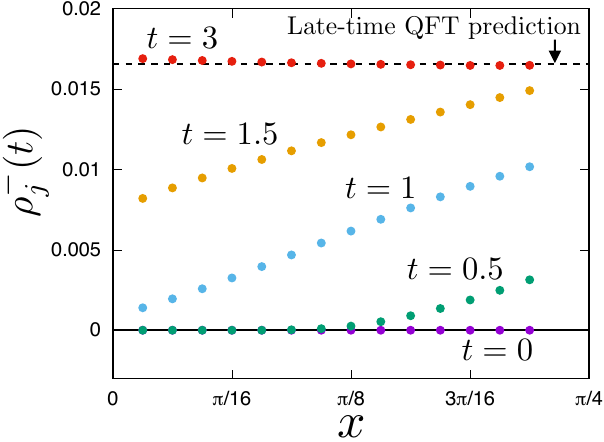}\label{Hmfig}
  }
  \subfigure[Norm of $\rho^\text{int}_j$]
 {\includegraphics[scale=0.65]{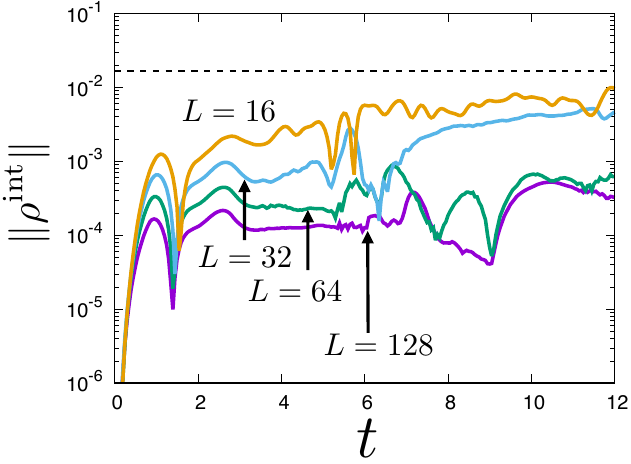}\label{Hint_abs}
  }
 \caption{Energy densities of (a) right-moving modes, (b) left-moving modes, and (c) the total energy density. The horizontal axis represents the spatial coordinate, and only the flat region $x < x_0$ is shown. The black dashed lines indicate the QFT predictions: Eq.~(\ref{Ehawking_pm}) for panels (a) and (b), and Eq.~(\ref{EhawkingTot}) for panel (c). Panel (d) shows the norm of $\rho^{\text{int}}_j$. For reference, Eq.~(\ref{Ehawking_pm}) is also shown as a black dashed line. The parameters are set to $\ell=\pi$, $x_0=\pi/4$, $\kappa=2$, and $\Delta t=2$. The system size is fixed at $L=64$ in panels (a)–(c), while in panel (d) it is varied as $L=16, 32, 64, 128$.
}
 \label{Hpmfig}
\end{figure}

In this section, we present a more detailed analysis of Hawking radiation in the spin system, complementing the results presented in the main text. While the main text focuses on the total energy density, Figs.~\ref{Hpmfig}(a) and (b) show the renormalized energy densities of the right- and left-moving modes, respectively, at several times.
The energy densities are shown only in the flat region satisfying $x_j<x_0-\varepsilon$. This restriction arises because the energy density at site $j$ involves operators at site $j+1$, so that the neighboring site to the right must also lie within the flat region.
The energy densities initially vanish and increase as the system evolves. After some time, they temporarily settle to values consistent with the QFT prediction~(\ref{Ehawking_pm}). Since Hawking radiation is initially generated as left-moving particles, the energy density of the left-moving modes reaches the QFT prediction earlier than that of the right-moving modes. The right-moving modes appear only after the Hawking radiation is reflected at the boundary at $x=0$, resulting in a delay of order $x_0$ relative to the left-moving modes. At later times, the energy densities eventually deviate from the QFT predictions because of lattice effects. Although this behavior is not shown in these panels, it becomes clearly visible in the subsequent figures.

To quantify the contribution from the interaction between the right- and left-moving modes, we define the norm of $\rho^\text{int}j$ as
\begin{equation}
\|\rho^\text{int}\|\equiv
\frac{1}{N}\sum_{x_{j}<x_0-\varepsilon} |\rho^\text{int}_j|\ ,
\end{equation}
where $N$ denotes the number of sites satisfying $x_j<x_0-\varepsilon$. Figure~\ref{Hint_abs} shows $\|\rho^\text{int}\|$ as a function of time. It remains sufficiently small compared with the energy density of Hawking radiation, indicated by the black dashed line corresponding to Eq.~(\ref{Ehawking_pm}). We also find that $|\rho^{\mathrm{int}}|\propto L^{-1}$, consistent with the fact that $\mathcal{H}_j^{\mathrm{int}}$ is subleading in the continuum limit.

\subsection{Dependence on the surface gravity and timescale of black hole formation}

The main text primarily presents results for $\kappa=2$ and $\Delta t=2$, in units where $\ell=\pi$. 

\begin{figure}
  \centering
\subfigure[$\kappa=1$, $\Delta t=2$]
 {\includegraphics[scale=0.7]{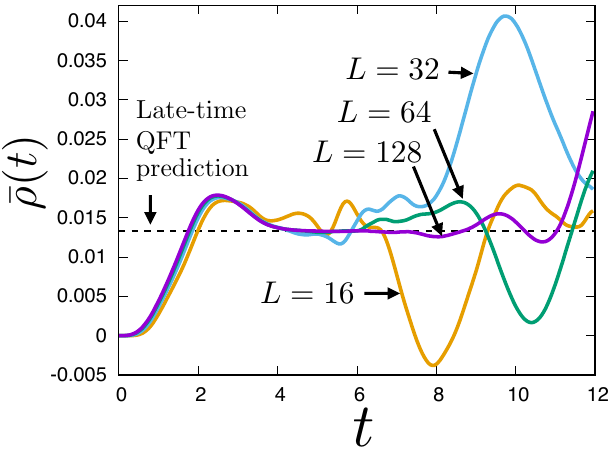}\label{kappa1}
  }
  \subfigure[$\kappa=3$, $\Delta t=2$]
 {\includegraphics[scale=0.7]{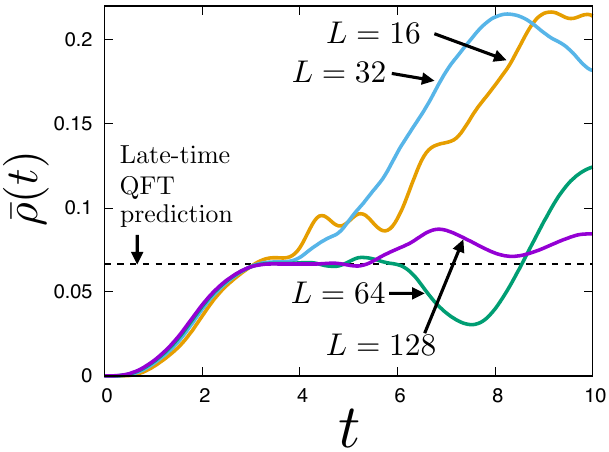}\label{kappa3}
  }
  \subfigure[$\kappa=2$, $\Delta t=0$]
 {\includegraphics[scale=0.7]{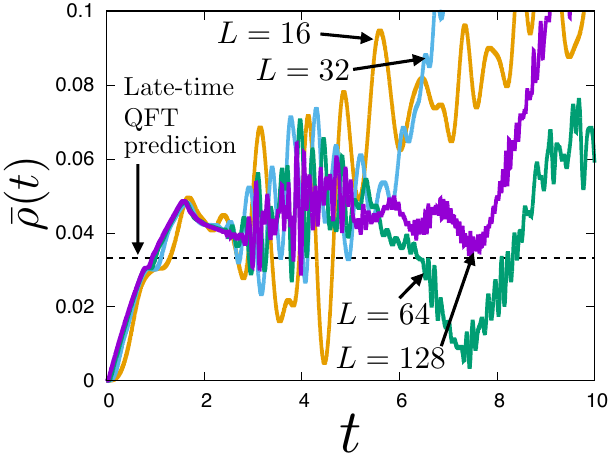}\label{dt0}
  }
  \subfigure[$\kappa=2$, $\Delta t=1$]
 {\includegraphics[scale=0.7]{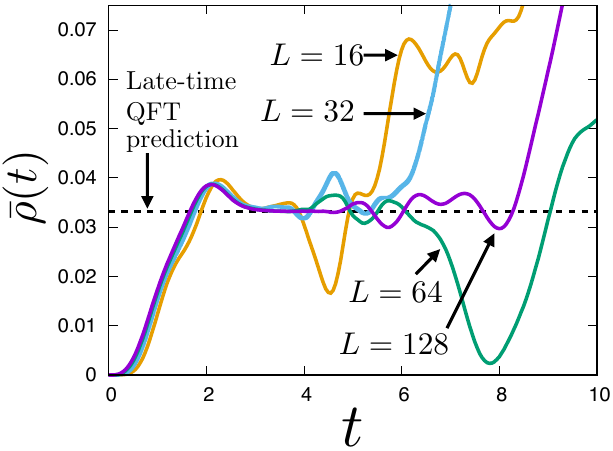}\label{dt1}
  }
 \caption{Panels (a)–(d) show the time dependence of the spatially averaged energy density, with $(\kappa,\Delta t)=(1,2)$, $(3,2)$, $(2,0)$, and $(2,1)$, respectively. Each panel shows curves for system sizes $L=16, 32, 64, 128$. Other parameters are set to $\ell=\pi$ and $x_0=\pi/4$.
}
 \label{Have_vary_k_dt}
\end{figure}

In the main text, we chose representative parameters to clearly demonstrate the energy density of Hawking radiation expected in the spin system. In this section, we extend the analysis to a wider range of parameters, focusing on the dependence on the surface gravity $\kappa$ and the timescale $\Delta t$ of the metric variation. Figure~\ref{Have_vary_k_dt} shows the spatially averaged energy density $\bar{\rho}(t)$ for $(\kappa,\Delta t)=(1,2)$, $(3,2)$, $(2,0)$, and $(2,1)$. Each panel shows the results for $L=16,32,64,128$.
We first consider panels (a), (b), and (d), for which the metric varies over a finite timescale ($\Delta t=1,2$). For all values $\kappa=1,2,3$, the spatially averaged energy density develops a plateau consistent with the QFT prediction. Although the plateau is less pronounced for $L=16$ and $32$ in panel (b), plateau-like behavior can still be observed over the short interval $3\lesssim t\lesssim4$.
Panel (c) corresponds to an instantaneous change of the metric ($\Delta t=0$). In this case, no clear plateau is observed, and small oscillations appear around the QFT prediction. Thus, although the energy density remains approximately centered around the QFT value, the agreement with QFT is less clear than for finite $\Delta t$. These oscillations are likely caused by the excitation of high-frequency modes induced by the instantaneous change of the metric, or equivalently, the Hamiltonian. These results suggest that, for a clear experimental observation of the Hawking-radiation energy density, the effective metric should be varied sufficiently smoothly.

\subsection{Convergence toward the continuum limit}
\begin{figure}
\begin{center}
\includegraphics[scale=0.6]{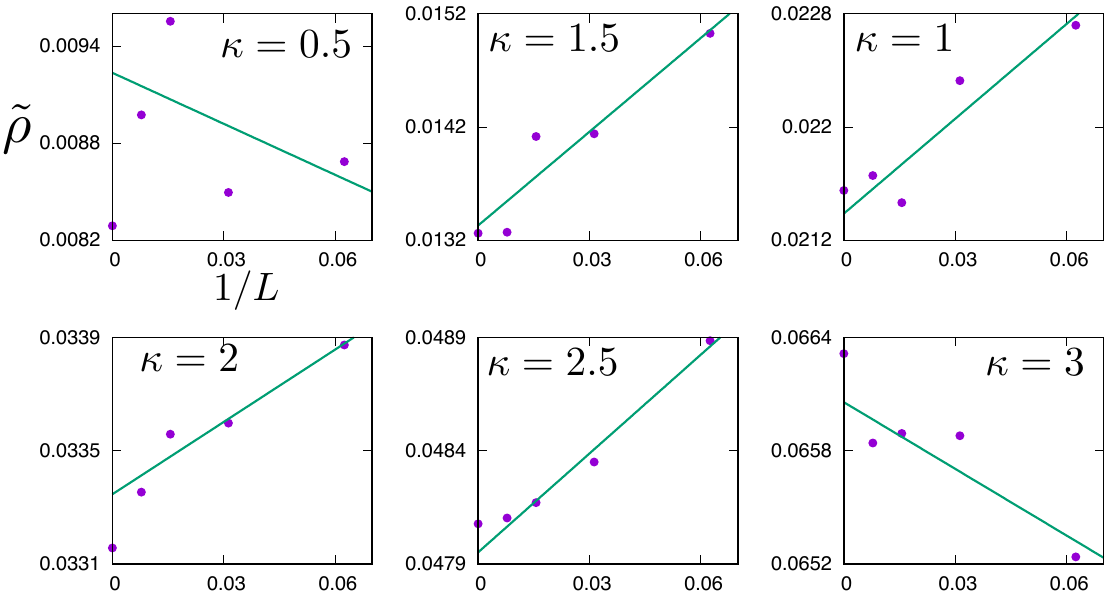}
\end{center}
\caption{Convergence check of the averaged energy density. The same set of parameters as those used in Fig.~\ref{Hplateau0} is adopted. The horizontal axis represents $1/L$, while the vertical axis shows $\tilde{\rho}$. The panels correspond to $\kappa = 0.5,\,1,\,1.5,\,2,\,2.5,$ and $3.0$, respectively. The green lines indicate linear fits to the numerical data points. The point at $1/L=0$ corresponds to the QFT value.
}
\label{conv_check}
\end{figure}

Here, we examine the convergence of the averaged energy density with increasing system size $L$. Figure~\ref{conv_check} displays the dependence of the averaged energy density on $1/L$. The results exhibit an approximate $1/L$ scaling, which is consistent with the expected $O(1/L)$ error associated with our finite-difference discretization. Although the convergence is less pronounced for $\kappa=0.5$, the extrapolated values remain close to the corresponding QFT prediction.

\subsection{Dynamics of the mode functions}
\label{nummode}

Here, we examine the behavior of the numerically obtained mode functions in the spin system. The mode functions are obtained by solving Eq.~(\ref{phieqspin}) using the fourth-order Runge--Kutta method. Throughout this subsection, we fix the parameters to $\ell=\pi$, $x_0=\pi/4$, $\kappa=2$, $\Delta t=2$, and $L=64$.

\begin{figure}
\begin{center}
\includegraphics[scale=0.5]{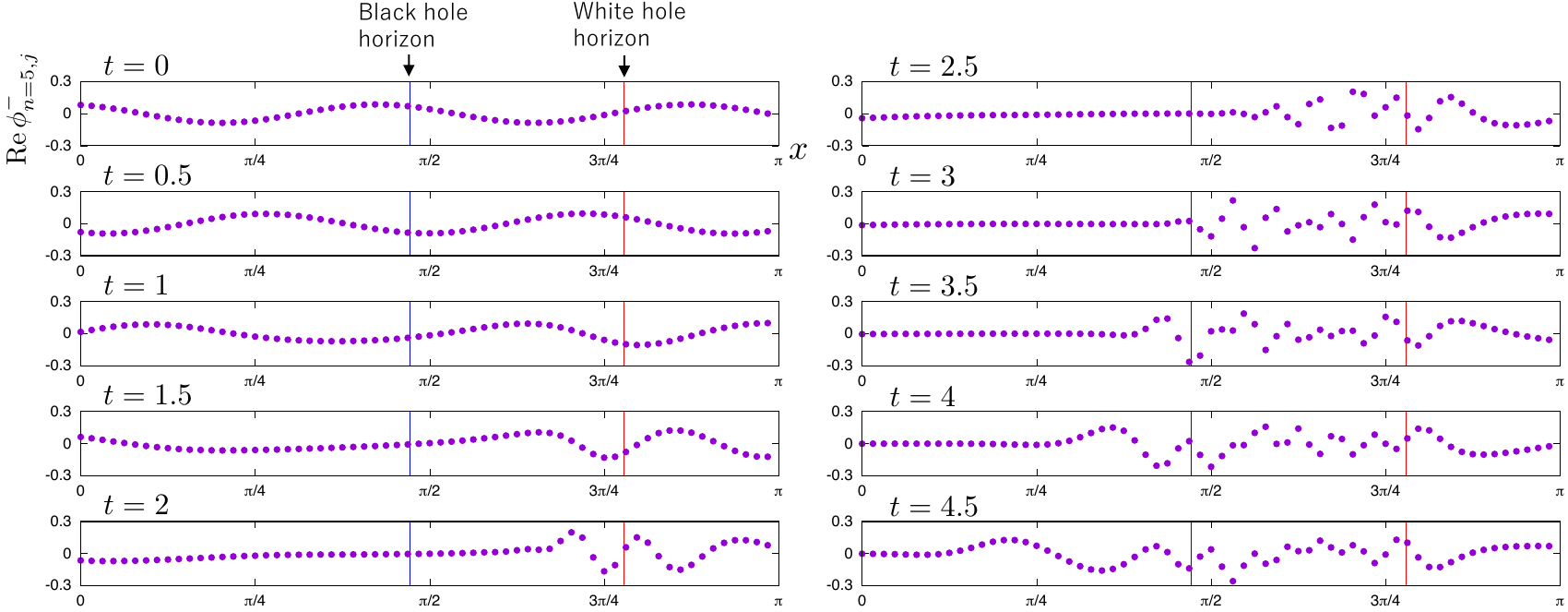}
\end{center}
\caption{Spacetime dependence of the mode function computed in the spin system. 
The real part of left-moving mode function $\phi^-_{n=5,j}$ with $n=5$ is shown. 
The parameters are set to $\ell=\pi$, $x_0=\pi/4$, $\kappa=2$, and $\Delta t=2$, and the system size is fixed at $L=64$. 
The blue and red vertical lines indicate the black-hole and white-hole horizons, respectively.
}
\label{modefunc5}
\end{figure}

Figure~\ref{modefunc5} shows the spacetime dependence of the mode functions. As a representative example, we plot the real part of $\phi^-_{n=5,j}$. At early times, the mode function evolves smoothly. Around $t\sim2.5$, however, rapidly varying spatial structures develop near the white-hole horizon. As time evolves further to $t\sim4$, these structures cross the black-hole horizon and propagate into the exterior region. In the continuum QFT, no signal can propagate from inside the black-hole horizon to the exterior. This behavior is therefore a discretization artifact, indicating that the discrete model no longer provides a good approximation to the continuum QFT at sufficiently late times. Its relation to the trans-Planckian problem is discussed in the main text.

\begin{figure}
\begin{center}
\includegraphics[scale=0.6]{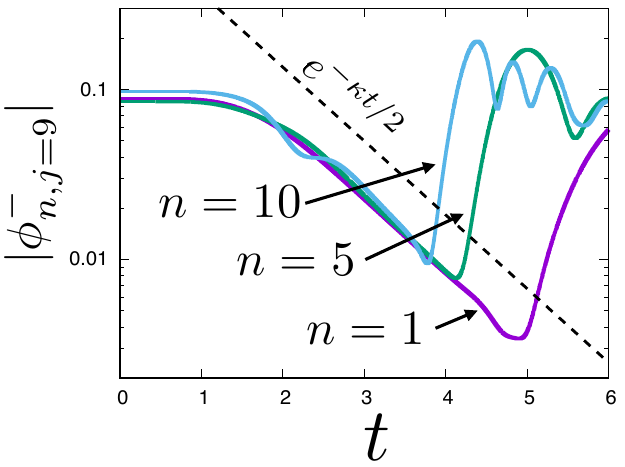}
\end{center}
\caption{Time dependence of the mode functions at a fixed spatial coordinate. 
The absolute values of the left-moving mode functions $\phi^-_{n,j}$ for $n=1,5,10$ are shown at $j=9$ ($x_9=\pi/8$). 
The black dashed line represents the late-time behavior predicted by QFT, $e^{-\kappa t/2}$. 
The parameters are set to $\ell=\pi$, $x_0=\pi/4$, $\kappa=2$, and $\Delta t=2$, and the system size is fixed at $L=64$.
}
\label{modefunc_fixed_j}
\end{figure}

Figure~\ref{modefunc_fixed_j} shows the time dependence of the mode functions at a fixed spatial coordinate, $x=\pi/8$ ($j=9$). We plot the absolute values of the left-moving mode functions $\phi^-_{n,j}$ for $n=1,5,10$. In the intermediate-time regime, the mode functions exhibit exponential decay, reflecting the absorption of excitations outside the horizon into the black hole. Indeed, the QFT result in Eq.~(\ref{phisollate}) predicts the late-time behavior $|\phi^-_{n,j}|\sim e^{-\kappa t/2}$, and the numerical results are consistent with this prediction over the intermediate-time regime. The deviation from the exponential decay at later times signals the breakdown of the discrete model as an approximation to the continuum theory, consistent with the behavior discussed above.

\bibliography{ref}